\documentclass[a4paper]{article}

\usepackage{amssymb,amsfonts,amsthm}
\usepackage{graphicx}
\usepackage{subfigure}
\usepackage{amssymb}
\usepackage{amsthm}
\usepackage{amsmath}
\usepackage{bm}             
\usepackage[mathscr]{eucal}

\usepackage{cancel}

\usepackage{psfrag}

\usepackage{wasysym}

\usepackage{color}

\newcommand{\hSigma}{\bar{\Sigma}}
\newcommand{\hN}{\bar{N}}
\newcommand{\hH}{\bar{H}}
\newcommand{\hF}{\bar{F}}
\newcommand{\hq}{\bar{q}}
\newcommand{\hS}{\bar{S}}
\newcommand{\hOmega}{\bar{\Omega}}
\newcommand{\htau}{\bar{\tau}_-}
\newcommand{\hT}{\bar{T}_-}
\newcommand{\hsigma}{\bar{\sigma}_-}

\newcommand{\BI}{\bm{B}_{\,\mathrm{I}}}
\newcommand{\BII}{\bm{B}_{\mathrm{II}}}
\newcommand{\BVI}{\bm{B}_{\mathrm{VI}_0}}
\newcommand{\BVII}{\bm{B}_{\mathrm{VII}_0}}
\newcommand{\BVIII}{\bm{B}_{\mathrm{VIII}}}
\newcommand{\BIX}{\bm{B}_{\mathrm{IX}}}

\newcommand{\overlineBII}{\overline{\bm{B}}_{\mathrm{II}}}
\newcommand{\overlineBVII}{\overline{\bm{B}}_{\mathrm{VII}_0}}
\newcommand{\overlineBIX}{\overline{\bm{B}}_{\mathrm{IX}}}

\newcommand{\weg}{\:\,}

\newcommand{\bom}{\bm{\omega}}

\newcommand{\diag}{\mathop{\mathrm{diag}}}

\newcommand{\spur}{\mathop{\mathrm{tr}}}
\newcommand{\textfrac}[2]{{\textstyle \frac{#1}{#2}}}

\theoremstyle{plain}
\newtheorem{theorem}{Theorem}[section]
\newtheorem{corollary}[theorem]{Corollary}

\newtheorem{lemma}[theorem]{Lemma}
\newtheorem*{definition}{Definition}

\theoremstyle{remark}

\newtheorem*{remark}{Remark}
\newcommand{\sfrac}[2]{{\textstyle{\frac{#1}{#2}}}}

\begin{document}


\title{\bf A new proof of the Bianchi type~IX attractor theorem}

\author{ \\
{\Large\sc J.\ Mark Heinzle}\thanks{Electronic address:
{\tt mark.heinzle@univie.ac.at}} \\[1ex]
Gravitational Physics, Faculty of Physics, \\
University of Vienna, A-1090 Vienna, Austria \\
and \\
Mittag-Leffler Institute of the Royal Swedish Academy of Sciences\\
S-18260 Djursholm, Sweden
\and \\
{\Large\sc Claes Uggla}\thanks{Electronic address:
{\tt claes.uggla@kau.se}} \\[1ex]
Department of Physics, \\
University of Karlstad, S-651 88 Karlstad, Sweden \\[2ex] }

\date{}
\maketitle

\begin{abstract}

We consider the dynamics towards the initial singularity of
Bianchi type~IX vacuum and orthogonal perfect fluid models with
a linear equation of state. The `Bianchi type~IX attractor
theorem' states that the past asymptotic behavior of generic
type~IX solutions is governed by Bianchi type~I and~II vacuum
states (Mixmaster attractor). 
We give a comparatively short and 
self-contained new proof of this theorem. The proof we give is
interesting in itself, but more importantly it illustrates and
emphasizes that type IX is special, and to some extent
misleading when one considers the broader context of generic
models without symmetries.

\end{abstract}

\newpage

\section{Introduction}\label{intro}

Based on work reviewed and developed in~\cite{waiell97} and by
Rendall~\cite{ren97}, Ringstr\"om succeeded, 
some eight, nine years ago,
to produce the first major theorem
about asymptotic dynamics
of type~IX vacuum and orthogonal perfect fluid
models~\cite{rin00,rin01}:  The Bianchi type~IX attractor
theorem, which
states that the past asymptotic behavior of generic
type~IX solutions is governed by Bianchi type~I and~II vacuum
states (which constitute the Mixmaster attractor); 
see Theorem~\ref{rinthm}.
In this paper we provide a comparatively short and
simple new proof of this theorem.
Our proof 
rests on three cornerstones:
\begin{itemize}
\item[(i)] We introduce new bounded variables that yield a relatively
    compact state space, which eliminates several of the
    complications associated with the unbounded variables
    used by Ringstr\"om.
\item[(ii)] We make systematic use of the Lie contraction
    hierarchy of invariant subsets admitted by Bianchi type
    IX where monotone functions restrict the asymptotic
    dynamics to boundaries of boundaries.
\item[(iii)] We systematically employ methods, arguments, and
    results from the theory of dynamical systems, see
    e.g.~\cite{waiell97,perko,cra91,col03} and references
    therein.
\end{itemize}

However, to find a more succinct argument is not our primary
motivation to re-investigate Bianchi type IX. Our proof
demonstrates that Bianchi type~IX is special in comparison with
the other oscillatory Bianchi models, i.e.,
type~$\mathrm{VI}_{-1/9}$ and type~VIII. This special nature of
Bianchi type~IX is associated with the geometric condition that
the three structure constants have the same sign, which results
in an extraordinary simplification of the
problem (which is easily overlooked) 
and thus makes the treatment of Bianchi type~IX models
relatively straightforward. The fact that Bianchi type~IX is
rather special has broad ramifications for our understanding of
generic spacelike singularities~\cite{uggetal03,heietal07}:
Bianchi type~IX is probably not quite as good a role model for
the asymptotic behavior of generic inhomogeneous models as is
commonly asserted.

Ringstr\"om's Bianchi type~IX attractor theorem is a remarkable
theorem. However, it is imperative to point out that this
theorem has limited implications, e.g., nothing is rigorously
known about dynamical chaotic properties (although there are
good grounds for beliefs). All claims about chaos in Einstein's
equations rest on the (plausible) belief that the asymptotic
dynamics of Einstein's equations is reflected by a discrete map
(the Kasner map), but this is far from evident and has not been
proved so far. These and related issues are discussed in the
paper `Mixmaster: Fact and Belief'~\cite{heiuggpreprint}. The
present paper, however, concentrates on rigorous results.

This paper is essentially self-contained. 
In Section~\ref{basic} we begin with a brief discussion of the
dynamical systems approach, where we establish the connection
with the metric approach. We briefly introduce
Hubble-normalized variables, but we concentrate on a set of new
bounded variables that yield a relatively compact state 
space---this
is the first cornerstone of our analysis.
In Section~\ref{subsets} we discuss the levels of the Bianchi
type~IX Lie contraction hierarchy: Apart from reviewing results
on the Bianchi type~I and type~II subsets, we present a
thorough analysis of Bianchi type~$\mathrm{VII}_0$. The proofs
we give are novel and, in particular, independent of results on
Bianchi type~IX---this is the second cornerstone of our
analysis. In Section~\ref{nongeneric} we present the results of
the local analysis of the fixed points of the dynamical system
and discuss
non-generic asymptotically self-similar behavior.
Section~\ref{newproof} is the core of this paper: We present a
new and succinct proof of the Bianchi type~IX attractor theorem
(which is stated as Theorem~\ref{rinthm}). The proof is
considerably shorter than the proof given by
Ringstr\"om~\cite{rin01} (which is in turn based on results
in~\cite{waiell97,ren97,rin00}). In the proof we systematically
employ methods from the theory of dynamical systems---the third
cornerstone of our analysis. In Section~\ref{consequences} we
state and prove a number of consequences of
Theorem~\ref{rinthm}. These results follow relatively easily
from Theorem~\ref{rinthm} when combined with the knowledge of
the flow on the Mixmaster attractor (which is the union of the
type~I and~II subsets). We conclude in Section~\ref{concl} with
a discussion of the main themes of this paper and we put
Bianchi type~IX in a broader context; in particular we
emphasize that the present Bianchi type~IX models are too
special in some respects to serve as good role models for
generic spacelike singularities. Throughout this paper we use
units so that $c=1$ and $8\pi G=1$, where $c$ is the speed of
light and $G$ the gravitational constant.

\section{Basic equations}
\label{basic}

Consider a vacuum or orthogonal perfect fluid spatially
homogeneous (SH) Bianchi type~IX model (i.e., the fluid
4-velocity is assumed to be orthogonal to the SH symmetry
surfaces). It is well known that there exists
a symmetry-adapted (co-)frame
$\{\hat{\bom}^1,\hat{\bom}^2,\hat{\bom}^3\}$ such that the
metric for these models takes the form
\begin{subequations}\label{bixform}
\begin{align}
\label{threemetric} & {}^4\mathbf{g}  = -dt\otimes dt +
g_{11}(t)\:\hat{\bom}^1\otimes \hat{\bom}^1 +
g_{22}(t)\:\hat{\bom}^2\otimes \hat{\bom}^2 +
g_{33}(t)\:\hat{\bom}^3\otimes \hat{\bom}^3\:,\\[0,5ex]
\label{structconst}
&\text{where}\quad
d\hat{\bom}^1  =  -\hat{n}_1 \, \hat{\bom}^2\wedge \hat{\bom}^3\:,\quad
d\hat{\bom}^2  =  -\hat{n}_2 \, \hat{\bom}^3\wedge \hat{\bom}^1\:,\quad
d\hat{\bom}^3  =  -\hat{n}_3 \, \hat{\bom}^1\wedge \hat{\bom}^2\:,
\end{align}
\end{subequations}
and where $\hat{n}_\alpha = {+1}$ $\forall \alpha$; see
e.g.~\cite{waiell97,rin01} and references therein.

\begin{remark}
Class~A Bianchi models of different Bianchi types are
characterized by different structure constants
$\hat{n}_\alpha$; the different cases are listed in
Table~\ref{classAmodels}.
\end{remark}

Let
\begin{equation}
\label{nsandmetric}
n_1(t) := \hat{n}_1 \,\frac{g_{11}}{\sqrt{\det g}} \:, \quad
n_2(t) := \hat{n}_2 \,\frac{g_{22}}{\sqrt{\det g}} \:,\quad
n_3(t) := \hat{n}_3 \,\frac{g_{33}}{\sqrt{\det g}} \:,
\end{equation}
where $\det g = g_{11} g_{22} g_{33}$. Furthermore, define
\begin{equation}
\theta = -\spur k\qquad\text{and}\qquad \sigma^\alpha_{\weg \beta} =
-k^{\alpha}_{\weg \beta} + \sfrac{1}{3} \spur k \:\delta^\alpha_{\weg \beta}=
\diag(\sigma_1,\sigma_2,\sigma_3) \quad \left(\Rightarrow
\sum\nolimits_\alpha \sigma_\alpha = 0\right)\:,
\end{equation}
where $k_{\alpha\beta}$ denotes the second fundamental form
associated with~\eqref{bixform} of the SH hypersurfaces
$t=\mathrm{const}$. The quantities $\theta$ and
$\sigma_{\alpha\beta}$ can be interpreted as the expansion and
the shear, respectively, of the normal congruence of the SH
hypersurfaces. The spatial volume density changes according to
according to $d\sqrt{\det g}/d t = \theta \sqrt{\det g}$. In a
cosmological context it is customary to replace $\theta$ by the
Hubble variable $H=\theta/3 = -\spur k/3$. Evidently, for the
present Bianchi type~IX models there is a one-to-one
correspondence between the `orthonormal frame variables' $(H,
\sigma_\alpha, n_\alpha)$ (with $\sum_\alpha \sigma_\alpha =0$)
and $(g_{\alpha\beta}, k_{\alpha\beta})$.

\begin{remark}
The orthonormal frame variables $(H,\sigma_\alpha, n_\alpha)$
can be used to describe any model of the family of Bianchi class~A
models; see Table~\ref{classAmodels}. In Bianchi type~VIII and~IX,
the metric is obtained from $(n_1,n_2,n_3)$
by~\eqref{nsandmetric}; for the lower Bianchi types, the other frame
variables, i.e.,  $(H, \sigma_\alpha)$, are needed as well to reconstruct the
metric; see~\cite{janugg99} for a group theoretical approach.
\end{remark}

\begin{table}
\begin{center}
\begin{tabular}{|c|ccc|}
\hline Bianchi type & $n_\alpha$ &  $n_\beta$ & $n_\gamma$ \\ \hline
I & $0$ & $0$ & $0$ \\
II & $0$ & $0$ & $+$ \\
$\mathrm{VI}_0$ & $0$ & $-$ & $+$ \\
$\mathrm{VII}_0$ & $0$& $+$ & $+$ \\
$\mathrm{VIII}$ & $-$& $+$& $+$ \\
$\mathrm{IX}$ & $+$& $+$& $+$ \\\hline
\end{tabular}
\caption{The class A Bianchi types are characterized by
different signs of the variables $(n_\alpha, n_\beta,
n_\gamma)$, where $(\alpha\beta\gamma)$ is any permutation of
$(123)$. In addition to the above representations there exist
equivalent representations associated with an overall change of
sign of the variables; e.g., another type IX representation is
$(---)$.} \label{classAmodels}
\end{center}
\end{table}

In the perfect fluid case we assume an orthogonal perfect fluid
with density $\rho$ and pressure $p$ that satisfy a linear
equation of state $p = w \rho$. We require the energy
conditions (weak/strong/dominant) to
hold, hence $\rho > 0$ and%
\footnote{Note, however, that the well-posedness of
  the Einstein equations (for solutions without symmetry)
  has been questioned in the case $-1/3<w<0$, see~\cite{friren00}.}
\begin{equation}\label{wassum}
-\sfrac{1}{3} < w < 1 \:;
\end{equation}
we exclude the special cases $w=-\textfrac{1}{3}$ and $w=1$, where the energy
conditions are only marginally satisfied.%
\footnote{The case $w=1$ is
  known as the stiff fluid case, for which the speed of sound is equal to
  the speed of light. The asymptotic dynamics of stiff fluid solutions is
  simpler than the oscillatory behavior characterizing the models with
  range $-\sfrac{1}{3} < w < 1$, and well understood~\cite{rin01,andren01}.
  (In the terminology introduced below,
  the stiff fluid models are asymptotically self-similar.)
  We will therefore refrain from discussing the stiff fluid case in this paper.}%

%
%

\subsection{Hubble-normalized variables and equations}

In the Hubble-normalized
dynamical systems approach
we define dimensionless
orthonormal frame variables according to
\begin{equation}\label{Hnormvars}
(\Sigma_\alpha, N_\alpha) = (\sigma_\alpha, n_\alpha)/H\:, \qquad
\Omega = \rho/(3H^2)\:.
\end{equation}

\begin{remark}
For all class A models except type~IX
the Gauss constraint guarantees that $H$ remains positive if it
is positive initially. In Bianchi type~IX, however, it is known from
a theorem by Lin and Wald~\cite{linwal89} that all type~IX vacuum
and orthogonal perfect fluid models with $w\geq 0$ first expand
($H>0$), reach a point of maximum expansion ($H=0$), and then
recollapse ($H<0$).%
\footnote{In the locally rotationally symmetric case it has been
  proved that
  the range of $w$ can be
  extended to $w>-\frac{1}{3}$, see~\cite{heietal05}. There are
  good reasons to believe that the assumption of local rotational symmetry
  is superfluous, but this has not been established yet.}
Therefore, although the variable transformation~\eqref{Hnormvars} breaks down at the
point of maximum expansion, the
variables $(\Sigma_\alpha, N_\alpha)$ correctly describe the
dynamics in the expanding phase.
\end{remark}

Since the past singularity is of particular interest in our
considerations, it is convenient to choose the \textit{time
direction towards the past singularity}.
We introduce a new dimensionless time variable $\tau_-$ according to
\begin{equation}\label{newtime}
d\tau_-/d t \, =\, {-}H\:.
\end{equation}

The Einstein fields equations can be reformulated in terms of
the Hubble variable $H$ and the Hubble-normalized variables~\cite{waiell97}.
Since $H$ is the only variable that carries dimension,
the equation for $H$,
\begin{equation}\label{H}
d H/d \tau_- = (1+q)H\:,
\end{equation}
decouples from the equations for $\Sigma_\alpha$ and
$N_\alpha$, which are given by the system (as follows
from~\cite{waiell97})
\begin{subequations}\label{IXeq}
\begin{align}
\label{sig}
d \Sigma_\alpha/d \tau_- & =  (2-q)\Sigma_\alpha  + {}^{3}\!S_\alpha\:, \\[0.5ex]
\label{n}
d N_\alpha/d \tau_- & =  -(q+2\Sigma_\alpha)\,N_\alpha
\qquad\qquad\qquad \text{(no sum over $\alpha$)}\:,
\end{align}
\end{subequations}
where
\begin{subequations}
\begin{alignat}{2}
\label{q}
q & = 2\Sigma^2 + \sfrac{1}{2}(1+3w)\Omega \:,
&  & \Sigma^2 = \sfrac{1}{6}(\Sigma_1^2 + \Sigma_2^2 + \Sigma_3^2)\:,\\
\label{threecurv}
\text{and}\quad  {}^{3}\!S_\alpha  & = \sfrac{1}{3}\left[ N_\alpha(2N_\alpha
- N_\beta - N_\gamma) - (N_\beta - N_\gamma)^2 \right],  & \quad &
(\alpha\beta\gamma) \in \left\{(123),(231),(312)\right \}\:.
\end{alignat}
\end{subequations}
It is straightforward to show that
$\rho \propto \exp\left(3[1+w]\tau_-\right)$
and $\tau_-\rightarrow\infty$ towards the past singularity, so that
$\rho\rightarrow\infty$ in this limit; see~\cite{heiuggpreprint}.


Apart from the trivial constraint
$\Sigma_1 + \Sigma_2 + \Sigma_3 = 0$,%
\footnote{It is common to globally solve $\Sigma_1 + \Sigma_2 + \Sigma_3 = 0$
  by introducing new variables according to
  $\Sigma_1 = -2 \Sigma_+$, $\Sigma_2 =
  \Sigma_+ - \sqrt{3} \Sigma_-$, $\Sigma_3 = \Sigma_+ + \sqrt{3} \Sigma_-$,
  which yields $\Sigma^2 = \Sigma_+^2 + \Sigma_-^2$. However, since this breaks the
  permutation symmetry of the three spatial axes (exhibited by type~IX models),
  we choose to retain the variables $\Sigma_1$, $\Sigma_2$, $\Sigma_3$.}
there exists the Gauss constraint
%
%
\begin{equation}
\label{gauss}
\Sigma^2 +
\sfrac{1}{12} \Big[ N_1^2 + N_2^2 + N_ 3^2 -
2 \big( N_1N_2 + N_2 N_3 + N_3N_1 \big) \Big] = 1 - \Omega \leq 1 \:.
\end{equation}
Since this constraint can be used to globally solve for
$\Omega$, the reduced state space is given as the space of all
$(\Sigma_1,\Sigma_2,\Sigma_3)$ and $(N_1,N_2,N_3)$
satisfying~\eqref{gauss} and $\Sigma_1 + \Sigma_2 + \Sigma_3 =
0$. The Gauss constraint~\eqref{gauss} reveals a serious
disadvantage of the Hubble-normalized variables: The range of
the variables is unbounded, i.e., the state space is not
relatively compact.


\subsection{Bounded variables} 
\label{domvarseqs}

The preparatory step in our approach to Ringstr\"om's Bianchi
type~IX attractor theorem is to reformulate the Einstein
equations for class~A Bianchi models in terms of variables that
span a relatively compact state space. (The aim is to avoid the
problems that are caused by the unboundedness of the
Hubble-normalized variables in the dynamical
system~\eqref{IXeq}; note, however, that the key arguments in
our proof, e.g., Lemma~\ref{noVII}, are independent of the
choice of variables.) Define
\begin{equation}\label{Ddef}
D := \sqrt{H^2 + \sfrac{1}{6} (n_1 n_2 + n_1 n_3 + n_2 n_3)}\:.
\end{equation}
The Gauss constraint reads $3 D^2 = \rho +
\frac{1}{2}(\sigma_1^2 + \sigma_2^2 + \sigma_3^2) +
\frac{1}{4}(n_1^2 + n_2^2 +n_3^2)$, hence $D > 0$. This makes
it possible to introduce variables that are normalized w.r.t.\
the `dominant' variable $D$ instead of $H$, i.e.,
\begin{equation}\label{domvars}
(\hSigma_\alpha,\hN_\alpha, \hH) =
(\sigma_\alpha,n_\alpha, H)/D\:,\qquad\quad \hOmega = \rho/(3 D^2)\:.
\end{equation}
By construction, $\hSigma_1 + \hSigma_2 + \hSigma_3 = 0$.

\begin{remark}
Contrary to the Hubble-normalized variables, the new
variables~\eqref{domvars} are globally well-defined.
\end{remark}

We choose a dimensionless time variable $\htau$, which is defined by
\begin{equation}\label{domtime}
\frac{d\htau}{d t} = -D\:;
\end{equation}
accordingly, $\htau$ is directed towards the past singularity.

The quantity $D$ decouples from the other equations for dimensional
reasons,
\begin{equation}\label{Deq}
\frac{d D}{d\htau} = \left[ (1 + \hq) \hH + \hF \right] D \:.
\end{equation}
The remaining dimensionless system of equations reads
\begin{subequations}\label{domsys}
\begin{align}
\frac{d\hH}{d\htau} & = \,\hq ( 1 - \hH^2) - \hF
\hH\:,\\
\frac{d\hSigma_\alpha}{d\htau} & =
\,\hSigma_\alpha \left[ (2-\hq) \hH - \hF \right] + {}^3\!\hS_\alpha \:,\\
\label{Nalpha}
\frac{d\hN_\alpha}{d\htau}  & = - \hN_\alpha
\left[ \hq \hH + 2 \hSigma_\alpha + \hF \right] \:,
\qquad\qquad\text{(no sum over $\alpha$)}
\end{align}
\end{subequations}
where
\begin{subequations}
\begin{alignat}{2}
\label{hq} \hq & = 2 \hSigma^2 + \sfrac{1}{2} (1 + 3 w)
\hOmega\:,
&  & \hSigma^2 = \sfrac{1}{6}(\hSigma_1^2 + \hSigma_2^2 + \hSigma_3^2)\:,\\
\hF & = \sfrac{1}{6} \left( \hN_1 \hN_2 \hSigma_3 +
\hN_1 \hSigma_2 \hN_3 + \hSigma_1 \hN_2 \hN_3 \right)\:, & &\\
{}^3\!\hS_\alpha & = \sfrac{1}{3} \left[ \hN_\alpha ( 2
\hN_\alpha - \hN_\beta - \hN_\gamma) - (\hN_\beta -
\hN_\gamma)^2 \right]\:,  & \quad & (\alpha\beta\gamma) \in
\left\{(123),(231),(312)\right \}\:.
\end{alignat}
\end{subequations}
The vacuum case is characterized by~$\hOmega = 0$ while
$\hOmega> 0$ in the fluid case. Note that the `deceleration
parameter' $\hq$ is non-negative.

In contrast to the system for the Hubble-normalized variables, there
exist two non-trivial constraints, the Gauss constraint and a constraint
resulting directly from~\eqref{Ddef},
\begin{subequations}\label{Dconstraints}
\begin{align}\label{Gausscon}
& \hSigma^2 + \sfrac{1}{12}
\left( \hN_1^2 + \hN_2^2 + \hN_3^2 \right) + \hOmega = 1\:, \\[1ex]
\label{Hconstraint} & \hH^2 + \sfrac{1}{6} \left( \hN_1 \hN_2  +
\hN_1  \hN_3 + \hN_2 \hN_3 \right) = 1\:.
\end{align}
\end{subequations}
Evidently, in Bianchi type~IX, the range of the new
variables~\eqref{domvars} is \textit{bounded}. This is true for
the entire class~A: We minimize the expression $\hN_1 \hN_2  +
\hN_1  \hN_3 + \hN_2 \hN_3$ appearing in~\eqref{Hconstraint}
under the condition $\hN_1^2 + \hN_2^2 + \hN_3^2   \leq 12$
resulting from~\eqref{Gausscon}; this leads to $\hN_1 \hN_2  +
\hN_1  \hN_3 + \hN_2 \hN_3 \geq -6$ and hence $\hH^2 \leq 2$
for all class A models; restriction to Bianchi type~IX yields
$\hH^2 < 1$, since $\hN_\alpha > 0$ $\forall \alpha$.
Therefore, the system~\eqref{domsys} is a system on a
relatively compact state space for the entire class~A.

The dimensionless state space of the Bianchi type~IX orthogonal
perfect fluid models with a linear equation of state is
5-dimensional while the state space of the vacuum models is
4-dimensional. The same is true for Bianchi type~VIII, while
the state spaces of the remaining class~A Bianchi models have
less degrees of freedom; see Table~\ref{Bianchistatespaces}.
Once the dynamics in the dimensionless state space is
understood, $D$ is obtained from a quadrature by
integrating~\eqref{Deq}, and subsequently the metric can be
reconstructed.
Note that we choose to use~\eqref{Gausscon} to solve for
$\hOmega$ when $\hOmega>0$; however, it is still useful to
consider the evolution equation
\begin{equation}
\label{hOmegaeq}
d\hOmega/d\htau =  -\hOmega \big[2 \hq \hH - (1 + 3w)\hH + 2 \hF]\:.
\end{equation}

\begin{remark}
Since $\hq$ is bounded as $\htau\rightarrow \infty$, the
asymptotics of $\hH$ can be bounded by exponential functions
from above and below. It follows that the equation $d t/d\htau
= -D^{-1}$ can be integrated to yield $t$ as a function of
$\htau$ such that $t\rightarrow 0$ as $\htau\rightarrow
{+}\infty$.
\end{remark}

In all Bianchi types except type~IX the constraints force $\hH$ to
have a sign for all $\htau$, e.g., $\hH > 0$ (which entails
that these models are forever expanding%
\footnote{This excludes the
Bianchi type~I and type~$\mathrm{VII}_{0}$ representations
of Minkowski spacetime.}%
). In Bianchi type~IX the subset $\hH > 0$ is a future
invariant subset of the state space, which follows from the
inequality $d\hH/d\htau \geq 0$ on $\hH = 0$. (There is a close
relationship between this fact and the results
of~\cite{linwal89}.)

Let $(\alpha\beta\gamma)$ denote any permutation of $(123)$.
We define the Bianchi type~IX state space $\BIX$ as the
future-invariant set
\begin{equation}\label{BIXdef}
\BIX = \left \{\hN_\alpha > 0,\hN_\beta > 0, \hN_\gamma > 0, \hH > 0, \,
\hSigma_\alpha, \hSigma_\beta,\hSigma_\gamma, \,\hOmega \geq 0\right\}\:,
\end{equation}
where the variables are subject to the
constraints~\eqref{Dconstraints} and $\hSigma_1 + \hSigma_2 +
\hSigma_3 = 0$; the set $\BIX$ comprises both the vacuum subset
and the fluid subset, i.e., $\hOmega \geq 0$.

Setting one or more of these variables $(\hN_\alpha,\hN_\beta,\hN_\gamma)$
to zero (which corresponds to Lie
contractions~\cite{jan01}) yields invariant boundary subsets which
represent more special Bianchi types.
In this spirit, the boundary%
\footnote{In our context, $\partial\bm{B}_\ast$ denotes
  $\overline{\bm{B}}_\ast\backslash \bm{B}_\ast$;
  the word `boundary' is chosen in lack of better terminology.
  Accordingly, since $\BIX = \left\{\hN_\alpha > 0,\hN_\beta > 0,
  \hN_\gamma > 0, \hH > 0, \hOmega \geq 0\right\}$,
  $\partial\BIX$ is the set where one of the variables $(\hN_\alpha,
  \hN_\beta, \hN_\gamma)$ or $\hH$ is set to zero, while
  the vacuum subset $\left\{\hN_\alpha > 0,\hN_\beta > 0, \hN_\gamma > 0,
  \hH > 0, \hOmega = 0\right\}$ does not appear in $\partial\BIX$.
  This convention is adapted to the
  formulation of Lemma~\ref{monlemma}.}
$\partial\BIX$ of the Bianchi type~IX state space is given by
\begin{equation}\label{partialBnine}
\partial\BIX = \overline{\bm{H}}_0 \cup \overlineBVII \:,
\end{equation}
where
\begin{subequations}
\begin{align}
\bm{H}_0 & = \{\hN_\alpha > 0,\hN_\beta > 0, \hN_\gamma > 0, \hH =
0, \hOmega \geq 0\}\:,\\[0.5ex]
\label{BVIIdef}
\BVII & = \{\hN_\alpha > 0,\hN_\beta > 0, \hN_\gamma = 0, \hH >
0, \hOmega \geq 0\}\:, \qquad (\alpha\beta\gamma) \in \left\{(123), (231),(312)
\right\}\:.
\end{align}
\end{subequations}
Note that $\BVII$ denotes the collection of the three
equivalent Bianchi type~$\mathrm{VII}_0$ subspaces; if we want
to refer to one of the subsets in particular we use the
notation $\mathcal{B}_{\hN_\alpha \hN_\beta}$ for the set
$\{\hN_\alpha > 0,\hN_\beta > 0, \hN_\gamma = 0, \hH > 0\}$.
The notation is such that the subscript denotes the non-zero
variables among $\{\hN_\alpha, \hN_\beta, \hN_\gamma\}$;
accordingly, $\BVII = \mathcal{B}_{\hN_1 \hN_2} \cup
\mathcal{B}_{\hN_2 \hN_3} \cup \mathcal{B}_{\hN_3 \hN_1}$.

The constraints~\eqref{Dconstraints} imply that the boundary of
the set $\bm{H}_0$ consists of the three points
\begin{equation}
\mathrm{Z}_\alpha:\:\, \hH = 0\:,\;\hSigma_\alpha=0\:, \;
\hN_\alpha =0\:,\; \hN_\beta = \hN_\gamma = \sqrt{6}\:, \;
\hOmega = 0\:, \qquad (\alpha\beta\gamma) \in \left\{(123),
(231),(312) \right\}\,;
\end{equation}
the boundary of $\BVII$ is the union of the closures of the
Bianchi type~II subspaces, which are given by
$\mathcal{B}_{\hN_\alpha} = \{\hN_\alpha > 0,\hN_\beta = 0,
\hN_\gamma = 0, \hOmega \geq 0\}$, $\alpha = 1,2,3$,
collectively denoted by $\BII$, and the three points
$\mathrm{Z}_\alpha$, i.e.,
\begin{equation}
\partial \BVII = \overlineBII \cup \{\mathrm{Z}_1, \mathrm{Z}_2, \mathrm{Z}_3\}\:.
\end{equation}
For completeness, we note that $\BI$ is the set $\{\hN_\alpha =
0,\hN_\beta = 0, \hN_\gamma = 0, \hOmega \geq 0\}$; we have
$\partial \BII = \BI$. A Bianchi subset contraction diagram for
type IX is given in Figure~\ref{contraction}.

\begin{remark}
The constraints imply that $\hH = 1$ on $\BI \cup \BII$;
hence the dominant variables reduce to the standard
Hubble-normalized variables for type~I and~II.
\end{remark}

\begin{remark}
Each of the sets $\BI$, $\BII$, $\BVII$, $\BIX$ (and
$\bm{H}_0$) decomposes into a vacuum subset ($\hOmega = 0$) and
a fluid subset ($\hOmega > 0$). We refer to these subsets using
the superscripts ${}^{\mathrm{vac.}}$ and ${}^{\mathrm{fl.}}$.
\end{remark}

\begin{remark}
Note that there does not appear a subset $\BVI$ as a boundary
subset of $\BIX$. In the concluding remarks,
Section~\ref{concl}, we argue that this simple fact has
far-reaching consequences.
\end{remark}


%
\begin{table}
\begin{center}
\begin{tabular}{|ccc|c|c|}
\hline Type & Symbol & Range of $(\hN_\alpha, \hN_\beta,
\hN_\gamma)$ & State space properties & $\mathsf{D}$ \\ \hline & & & & \\[-1.8ex]
I & $\BI$ &
$\hN_\alpha =0$, $\hN_\beta = 0$,
$\hN_\gamma = 0$ &  $\hH \equiv 1$, $\hSigma^2 \leq 1$ & 2 \\[0.2ex]
II & $\BII$ &  $\hN_\alpha =0$, $\hN_\beta = 0$, $\hN_\gamma > 0$ &
$\hH\equiv 1$,
 $\hSigma^2 +\sfrac{1}{12} \hN_\gamma^2 \leq 1$ & 3\\[0.2ex]
$\mathrm{VI}_0$ & $\BVI$ & $\hN_\alpha = 0$, $\hN_\beta <0$, $\hN_\gamma > 0$
&  $\hH \in (0,2)$,
$\hSigma^2 +\sfrac{1}{12} [\hN_\beta -  \hN_\gamma]^2 \leq \hH^2$ & 4 \\[0.2ex]
$\mathrm{VII}_0$ & $\BVII$ &  $\hN_\alpha = 0$, $\hN_\beta > 0$, $\hN_\gamma > 0$
& $\hH \in (0,1)$,
$\hSigma^2 +\sfrac{1}{12} [\hN_\beta - \hN_\gamma]^2 \leq \hH^2$ & 4\\[0.2ex]
$\mathrm{VIII}$ & $\BVIII$ & $\hN_\alpha < 0$, $\hN_\beta > 0$, $\hN_\gamma > 0$
& $\hH \in (0,2)$, $\hSigma^2 <\hH^2$ & 5 \\[0.2ex]
$\mathrm{IX}$ & $\BIX^{\:\text{\tiny (\!$\ast$\!)}}$ & $\hN_\alpha > 0$, $\hN_\beta > 0$, $\hN_\gamma
> 0$ &  $\hH \in ({-1}, 1)$, $\hSigma^2 \not<\hH^2$
& 5 \\[0.2ex]\hline
\end{tabular}
\caption{The dimensionless state spaces associated with class~A
Bianchi models; here, $(\alpha\beta\gamma)$ is any permutation
of $(123)$. In addition to the above representations there
exist equivalent representations associated with an overall
change of sign of the variables $(\hN_1,\hN_2,\hN_3)$ or $\hH$.
The quantity $\mathsf{D}$ denotes the dimension of the state
space (in the fluid case); the dimensionality of the state
space in the vacuum cases is given by $\mathsf{D} - 1$.
The Bianchi type~IX state space decomposes into a future-invariant half
and a past-invariant half. By $\BIX$ we denote the future-invariant half,
i.e., $\hH > 0$; cf.~\eqref{BIXdef}.} \label{Bianchistatespaces}
\end{center}
\end{table}
%

\section{The boundaries of the state space: $\BI$, $\BII$, $\BVII$}
\label{subsets}

\begin{figure}[ht]
\psfrag{a}[cc][cc]{$\mathcal{B}_{\hN_1\hN_2\hN_3}$}
\psfrag{b}[cc][cc]{$\mathcal{B}_{\hN_1\hN_2}$}
\psfrag{c}[cc][cc]{$\mathcal{B}_{\hN_2\hN_3}$}
\psfrag{d}[cc][cc]{$\mathcal{B}_{\hN_1\hN_3}$}
\psfrag{e}[cc][cc]{$\mathcal{B}_{\hN_1}$}
\psfrag{f}[cc][cc]{$\mathcal{B}_{\hN_2}$}
\psfrag{g}[cc][cc]{$\mathcal{B}_{\hN_3}$}
\psfrag{h}[cc][cc]{$\mathcal{B}_\emptyset$}
\psfrag{i}[rc][cc]{$\BIX$}
\psfrag{j}[rc][cc]{$\BVII$}
\psfrag{k}[rc][cc]{$\BII$}
\psfrag{l}[rc][cc]{$\BI$} \centering{
  \includegraphics[height=0.45\textwidth]{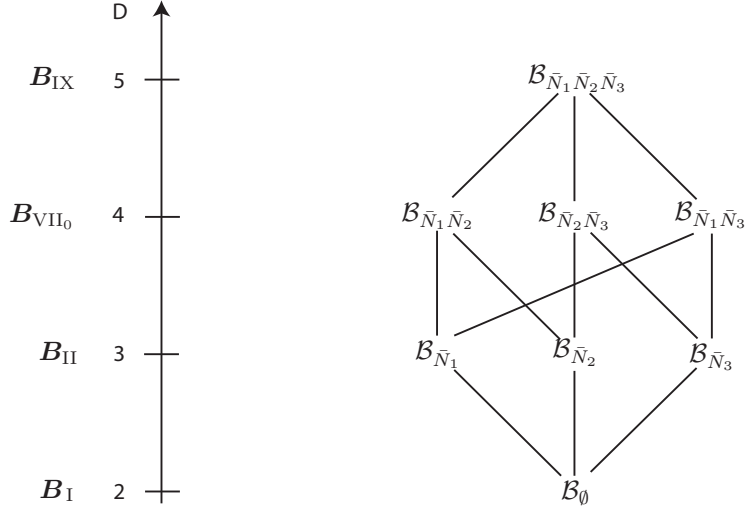}}
\caption{Subset contraction diagram for Bianchi type IX.
$\mathsf{D}$ denotes the dimension of the dimensionless state space
for the various models with an orthogonal perfect fluid with linear
equation of state; the associated vacuum subsets have one dimension
less; see also Table~\ref{Bianchistatespaces}. The notation is such
that the subscript of $\mathcal{B}_{\ast}$ denotes the non-zero
variables, e.g. $\mathcal{B}_{\hN_1\hN_2}$ denotes the type
$\mathrm{VII}_0$ subset with $\hN_1>0$, $\hN_2 >0$ and $\hN_3 = 0$.}
\label{contraction}
\end{figure}

In the analysis of a dynamical system that is defined on a
(relatively) compact state space, the analysis of the boundary
subsets is crucial. Often there exists a hierarchy of boundary
subsets that governs the asymptotic behavior of solutions, see
e.g.~\cite{heiugg06,calhei07} for examples how such hierarchies
can be exploited. In the case of Bianchi type~IX models, the
`Lie contraction hierarchy' of the boundaries is depicted in
Figure~\ref{contraction}.

%

\subsection{The Bianchi type I subset}

The Bianchi type~I subset $\BI$ is given by $\hN_1 = 0$, $\hN_2
=0$, $\hN_3 = 0$, hence $\hH=1$ and $\hOmega = 1 -\hSigma^2
\geq 0$. The \textit{vacuum subset}, $\hOmega = 0$, consists of
a circle of fixed points---\textit{the Kasner circle}
$\mathrm{K}^{\ocircle}$, which is characterized by
$\hSigma^2=1$. Each fixed point on $\mathrm{K}^{\ocircle}$
represents a Kasner solution (Kasner metric). There exist six
special points: $\mathrm{Q}_\alpha$ are given by
$(\hSigma_\alpha, \hSigma_\beta, \hSigma_\gamma) = ({-2},1,1)$;
the Taub points $\mathrm{T}_\alpha$ are given by
$(\hSigma_\alpha, \hSigma_\beta, \hSigma_\gamma) =
(2,{-}1,{-}1)$. The former are associated with locally
rotationally symmetric (LRS) solutions whose intrinsic geometry
is non-flat; the latter correspond to the flat LRS
solutions---the Taub representation of Minkowski spacetime.

The Bianchi type~I perfect fluid subset is the set
$1- \hOmega = \hSigma^2 < 1$. From~\eqref{hOmegaeq} it is
straightforward to deduce that there exists a central fixed
point, the \textit{Friedmann fixed point} $\mathrm{F}$, given
by $\hSigma_1=\hSigma_2=\hSigma_3=0$, which corresponds to the
isotropic Friedmann-Robertson-Walker (FRW) solution. Solutions
with $0<\hSigma^2 <1$ are given by radial straight lines
originating from $\mathrm{F}$ and ending at
$\mathrm{K}^\ocircle$. These results rely on the assumption $w
< 1$, see~\eqref{wassum}.

\subsection{The Bianchi type II subset}
\label{typeIIsubset}

The Bianchi type~II subset $\BII$ has three equivalent
representations: $\mathcal{B}_{\hN_1}$, $\mathcal{B}_{\hN_2}$,
$\mathcal{B}_{\hN_3}$. Let us consider
$\mathcal{B}_{\hN_\gamma}$, which is given by $\hN_\alpha =
\hN_\beta = 0$, $\hN_\gamma > 0$ (hence $\hH=1$). On
$\mathcal{B}_{\hN_\gamma}$, the Gauss constraint $\hSigma^2 +
\sfrac{1}{12} \hN_\gamma^2 + \hOmega =1$ can be used to replace
$\hN_\gamma$ by $\hOmega$ as a dependent variable. The
system~\eqref{IXeq} thus becomes
\begin{equation}\label{IIeq}
\frac{d\hSigma_{\alpha/\beta}}{d\htau} =
(2-q) \hSigma_{\alpha/\beta} +{}^3\!S_{\alpha/\beta} \,,
\quad
\frac{d\hSigma_\gamma}{d\htau} = (2-q) \hSigma_{\gamma} + {}^3\!S_{\gamma}\,,
\quad
\frac{d\hOmega}{d\htau} = -\hOmega \left[ 2 q -(1+3 w) \right]\,,
\end{equation}
where $q = 2 \hSigma^2 + \sfrac{1}{2} (1+3 w) \hOmega$ and
${}^3\!S_{\alpha/\beta} = -4 (1-\hSigma^2-\hOmega)$,
${}^3\!S_{\gamma} = 8 (1-\hSigma^2-\hOmega)$; we have $\hSigma^2 + \hOmega < 1$.

\begin{figure}[Ht]
\psfrag{t1}[rc][cc][0.9][0]{$\mathrm{T}_1$}
\psfrag{t2}[tl][tl][0.9][0]{$\mathrm{T}_2$}
\psfrag{t3}[bl][bl][0.9][0]{$\mathrm{T}_3$}
\psfrag{q1}[lc][cc][0.9][0]{$\mathrm{Q}_1$}
\psfrag{q2}[br][br][0.9][0]{$\mathrm{Q}_2$}
\psfrag{q3}[tr][tr][0.9][0]{$\mathrm{Q}_3$}
\centering
\includegraphics[width=0.25\textwidth]{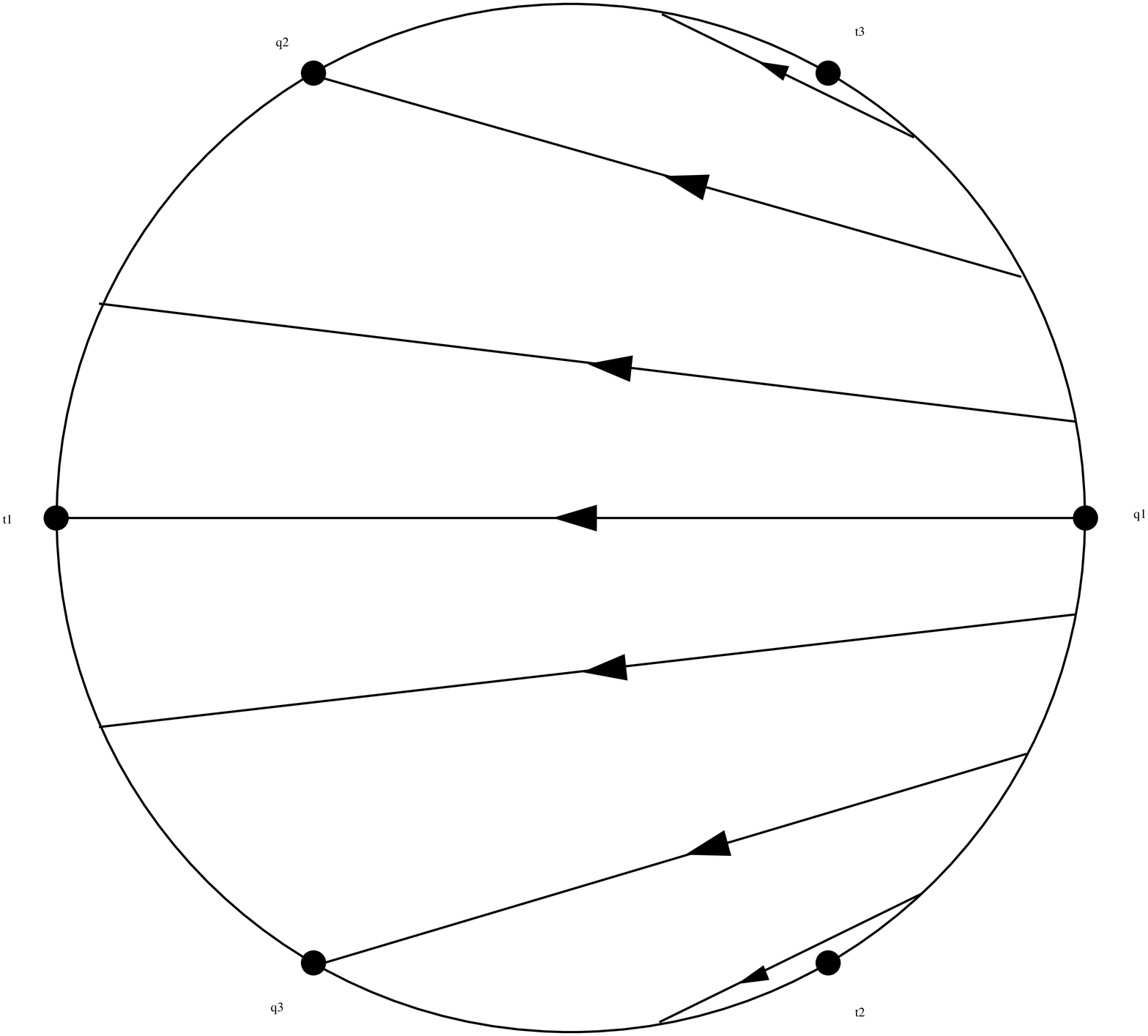}\qquad\qquad
\includegraphics[width=0.25\textwidth]{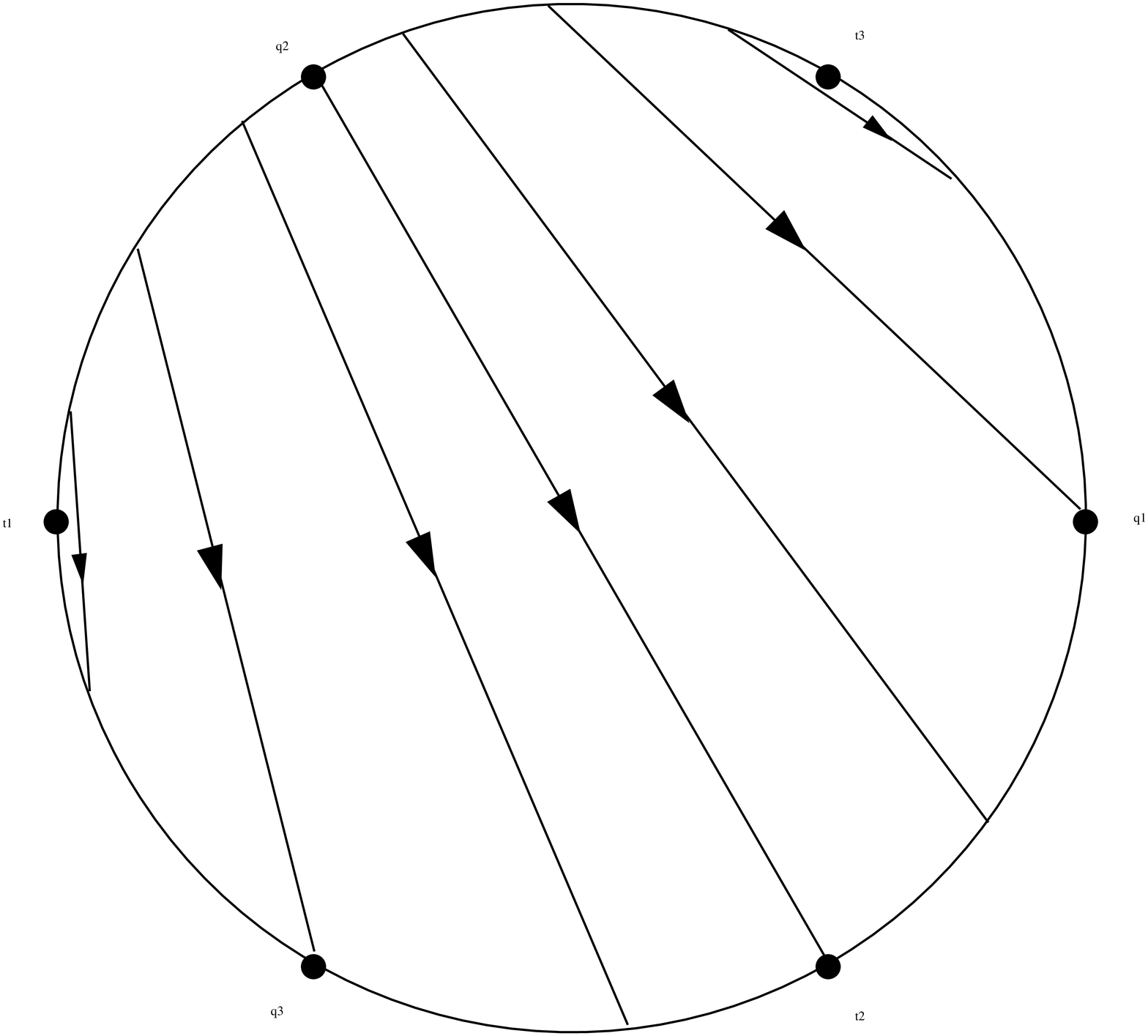}\qquad\qquad
\includegraphics[width=0.25\textwidth]{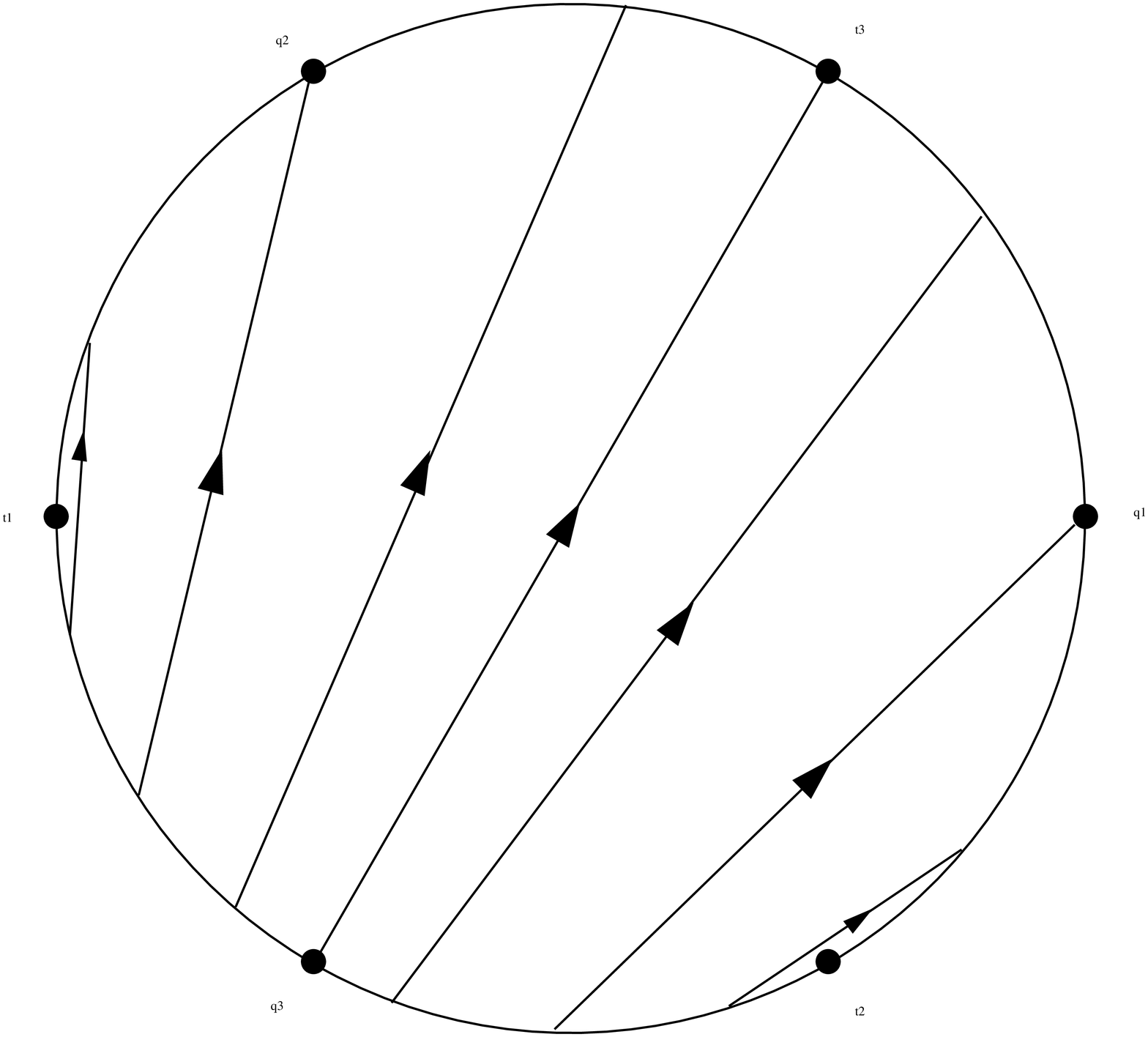}
\caption{Projection of the flow on the type II subsets
$\mathcal{B}_{\hN_1}^{\mathrm{vac.}}$, $\mathcal{B}_{\hN_2}^{\mathrm{vac.}}$, $\mathcal{B}_{\hN_3}^{\mathrm{vac.}}$
onto $(\hSigma_1, \hSigma_2,\hSigma_3)$-space.}
\label{alltypeII}
\end{figure}

Let us first consider the vacuum subset
$\mathcal{B}_{\hN_\gamma}^{\mathrm{vac.}}$, i.e., $\hOmega =
0$. There do not exist any fixed points on
$\mathcal{B}_{\hN_\gamma}^{\mathrm{vac.}}$, but the boundary
coincides with the Kasner circle $\mathrm{K}^\ocircle$. The
orbits of~\eqref{IIeq} form a family of straight lines in
$\mathcal{B}_{\hN_\gamma}^{\mathrm{vac.}}$; each orbit is a
heteroclinic orbit, since it connects two different fixed
points. If the initial point is $\mathrm{Q}_\gamma$, the final
point is $\mathrm{T}_\gamma$ (LRS orbit); the points
$\mathrm{T}_\alpha$ and $\mathrm{T}_\beta$ are not connected
with any other fixed point (they are `fixed points' under the
present `vacuum Bianchi type II map'), see
Figure~\ref{alltypeII}.

While there do not exist any fixed points in 
$\mathcal{B}_{\hN_\gamma}^{\mathrm{vac.}}$, there exists one
fixed point in $\mathcal{B}_{\hN_\gamma}^{\mathrm{fl.}}$ with
$\hOmega >0$, the \textit{Collins-Stewart fixed point}
$\mathrm{CS}_\gamma$, which corresponds to one representation
of the LRS solutions found by Collins and
Stewart~\cite{colste71}. $\mathrm{CS}_\gamma$ is given by
$(\hSigma_\alpha,\hSigma_\beta,\hSigma_\gamma) =
\sfrac{1}{8}(1+3 w) (1,1,-2)$ and $\hOmega = 1 - \sfrac{1}{16}
(1+3 w)$ (which yields $\hN_\gamma = \sfrac{3}{4}
\sqrt{1-w}\sqrt{1+3 w}\,$). The fixed point
$\mathrm{CS}_\gamma$ is the source for all orbits in
$\mathcal{B}_{\hN_\gamma}^{\mathrm{fl.}}$. For a detailed
discussion of these results see~\cite{waiell97}.

\subsection{Bianchi type $\mathbf{VII}_{\bm{0}}$: A new analysis}
\label{Bianchisevenzero}

A detailed analysis of the dynamics of Bianchi
type~$\mathrm{VII}_0$ models is essential for an understanding of
type~IX asymptotic dynamics. To underline the importance of
Bianchi type~$\mathrm{VII}_0$, we present a
new analysis of the global dynamics of type~$\mathrm{VII}_0$ solutions;
the proofs we give are novel and, in particular, independent
of results on Bianchi type~IX.
(The proof given in~\cite{rin01,rin03} relies on results on Bianchi type~IX.)

The dynamical system on $\BVII$ ($= \mathcal{B}_{\hN_1 \hN_2}
\cup \mathcal{B}_{\hN_2 \hN_3} \cup \mathcal{B}_{\hN_3 \hN_1}$)
does not admit fixed points with $\hOmega > 0$ (while, of
course, there exist fixed points on $\partial \BVII$). However,
each of the three vacuum subsets $\mathcal{B}_{\hN_\alpha
\hN_\beta}^{\,\mathrm{vac.}}$ contains a line of fixed points
with $\hN_\alpha = \hN_\beta >0$:
\begin{subequations}
\begin{equation}\label{TLdef}
\mathrm{TL}_\gamma =
\left\{ 0 < \hH < 1, \hSigma_\alpha =\hSigma_\beta =
-\hH , \hSigma_\gamma = 2 \hH,
\hN_\alpha = \hN_\beta = \sqrt{6} \sqrt{1-\hH^2}, \hN_\gamma =0,
\hOmega = 0  \right\}.
\end{equation}
These lines of fixed points (`Taub lines') connect the Taub points
$\mathrm{T}_\gamma$ on the Kasner circle with the fixed points
$\mathrm{Z}_\gamma$.
Each of the fixed points on $\mathrm{TL}_\gamma$
represents the Minkowski spacetime
in a Bianchi type $\mathrm{VII}_0$ LRS symmetry foliation;
this is in analogy to the Taub points $\mathrm{T}_\gamma$ themselves.
(Note that $\hSigma_\alpha =\hSigma_\beta = -\hH$,
$\hSigma_\gamma = 2 \hH$ corresponds to $\Sigma_\alpha =
\Sigma_\beta = -1$, $\Sigma_\gamma = 2$ in Hubble-normalized variables.
$\hN_\alpha = \hN_\beta \in (0,\sqrt{6})$ corresponds to
$N_\alpha = N_\beta \in (0,\infty)$.)
The second family of
LRS vacuum subsets are the three sets
\begin{equation}\label{QLdef}
\mathcal{QL}_\gamma =
\left\{ 0 < \hH < 1, \hSigma_\alpha =\hSigma_\beta = \hH , \hSigma_\gamma = -2 \hH,
\hN_\alpha = \hN_\beta = \sqrt{6} \sqrt{1-\hH^2}, \hN_\gamma =0, \hOmega = 0  \right\},
\end{equation}
\end{subequations}
which connect the exceptional points $\mathrm{Z}_\gamma$ with the
non-flat LRS fixed points $\mathrm{Q}_\gamma$.
These lines are not lines of fixed points,
since $d\hH/d\htau = 4 \hH^2 ( 1 - \hH^2)$, but heteroclinic orbits
$\mathrm{Z}_\gamma\rightarrow \mathrm{Q}_\gamma$.
The corresponding
solutions represent the non-flat LRS Kasner solutions in a Bianchi
type~$\mathrm{VII}_0$ symmetry foliation.

\textbf{The vacuum case} $\bm{\hOmega = 0}$. On the representative
$\mathcal{B}_{\hN_\alpha \hN_\beta}^{\,\mathrm{vac.}}$ of $\BVII^{\mathrm{vac.}}$
consider
the function $( 2 \hH - \hSigma_\gamma)$.
This function is non-negative, since $\hSigma_\gamma \in [-2 \hH,
2 \hH]$, which follows from the constraints when we use the relation
$12 \hSigma^2 = 3 \hSigma_\gamma^2 + (\hSigma_\alpha-\hSigma_\beta)^2$;
furthermore, $( 2 \hH - \hSigma_\gamma)$ is increasing on
$\overline{\mathcal{B}}_{\hN_\alpha \hN_\beta}^{\,\mathrm{vac.}}
\backslash \left( \overline{\mathrm{TL}}_\gamma \cup \mathrm{K}^\ocircle\right)$,
since
%
%
\begin{equation}\label{monvii0}
\frac{d}{d\htau} \left( 2 \hH - \hSigma_\gamma \right) =
\sfrac{1}{24} \left( 2 \hH - \hSigma_\gamma \right) \left[
(\hN_\alpha + \hN_\beta)^2  \left( 2 \hH - \hSigma_\gamma \right)
+ (\hN_\alpha - \hN_\beta)^2 \left( 2 \hH + \hSigma_\gamma\right) \right]\:.
\end{equation}
The monotonicity principle%
\footnote{The monotonicity
  principle~\cite{waiell97} gives information about the global
  asymptotic behavior of solutions of a dynamical system. If $M$
  is a ${\mathcal C}^1$ function on the state space $X$ that
  is strictly decreasing along orbits, then
  \begin{equation}
    \omega(x) \subseteq
    \{\xi \in \bar{X}\backslash X\:|\: \lim\limits_{\zeta\rightarrow \xi} M(\zeta) \neq
    \sup\limits_{X} M\}\:,\qquad
    \alpha(x) \subseteq
    \{\xi \in \bar{X}\backslash X\:|\:\lim\limits_{\zeta\rightarrow \xi} M(\zeta) \neq
    \inf\limits_{X} M\}\nonumber
  \end{equation}
  for all $x\in X$.}
implies that the $\alpha$-limit is a fixed point on
$\overline{\mathrm{TL}}_\gamma = \{\mathrm{T}_\gamma\} \cup
\mathrm{TL}_\gamma \cup \{\mathrm{Z}_\gamma\}$, since the fixed
points on the Kasner
circle 
are excluded as possible $\alpha$-limit points because
they are (transversally hyperbolic) saddles and sinks.

\begin{remark}
In addition, the monotonicity principle implies that the
$\omega$-limit of every orbit in $\mathcal{B}_{\hN_\alpha \hN_\beta}^{\,\mathrm{vac.}}$
(except for the fixed points on $\mathrm{TL}_\gamma$
themselves) is a fixed point on the Kasner circle (where
$\mathrm{T}_\gamma$ is excluded); clearly, only the
transversally hyperbolic sinks come into question.
\end{remark}

The results concerning the $\alpha$-limits of orbits
that are obtained from the monotonicity principle can
be strengthened considerably:

\begin{lemma}\label{sevenlemma}
The only type~$\mathrm{VII}_0$ vacuum
orbit that converges to $\mathrm{Z}_\gamma$ as $\htau
\rightarrow -\infty$ is the orbit $\mathcal{QL}_\gamma$; any other
orbit converges to one of the points on $\mathrm{TL}_\gamma$ as $\htau
\rightarrow -\infty$; conversely, each point on $\mathrm{TL}_\gamma$
is the $\alpha$-limit set for a one-parameter set of orbits.
\end{lemma}

The proof of these statements requires a more detailed analysis
of the dynamical system (in the neighborhood of $\mathrm{TL}_\gamma$,
in particular). The reader who is not interested in these details
might prefer to continue with the discussion of the Bianchi
type~$\mathrm{VII}_0$ fluid case below; note, however,
that the proof we give here is independent of
the proof given in~\cite{rin01,rin03} (and, in particular, it is
completely independent of results on Bianchi type~IX).

\begin{proof}
Let us introduce a set of variables that are adapted to the special features of
the Bianchi type~$\mathrm{VII}_0$ state space, using the Hubble-normalized
formulation as a starting point.%
\footnote{It follows from
  Section~\ref{basic} that $\Sigma_\delta = \hSigma_\delta/\hH$
  and $N_\delta = \hN_\delta/\hH$, $\delta=1,2,3$;
  in particular, $\hSigma_\gamma = 2 \hH \Leftrightarrow \Sigma_\gamma =2$;
  the time variables satisfy $d\tau_- = \hH d\htau$.}
Consider
$\mathcal{B}_{\hN_\alpha \hN_\beta}^{\,\mathrm{vac.}}$ and let
\begin{equation}
\Sigma_\alpha - \Sigma_\beta = 2 \sqrt{3}\, \sin\vartheta \cos 2\psi,\quad
\Sigma_\gamma = 2 \cos\vartheta,\quad
N_\alpha - N_\beta = \sqrt{12}\, \sin\vartheta \sin 2\psi ,\quad
N_\alpha N_\beta = 3 \zeta^2\,,
\end{equation}
where $(\vartheta,\psi,\zeta) \in [0,\pi] \times [0,\pi) \times (0,\infty)$.
The condition $\vartheta = \pi$ yields $\mathcal{QL}_\gamma$, while
$\vartheta = 0$ yields $\mathrm{TL}_\gamma$.
The constraint is automatically satisfied by this choice of variables;
the dynamical system takes the form
\begin{subequations}\label{polsys}
\begin{align}
\label{varthetaeq}
\frac{d\vartheta}{d\tau_-} & = (1-\cos\psi) (1-\cos\vartheta) \sin\vartheta \:,\\
\label{psieq}
\frac{d\psi}{d\tau_-} & = -4 \sin \vartheta
\sqrt{3 \Big(\frac{\zeta^2}{\sin^2\vartheta} + \frac{1- \cos\psi}{2}\Big)}
- 2\sin\psi\, (1 -\cos\vartheta) \:,\\
\label{zetaeq}
\frac{d\zeta}{d\tau_-} & = - \zeta \big[ 2 (1-\cos\vartheta) - \sin^2\vartheta\, ( 1 - \cos\psi) \big]\:.
\end{align}
\end{subequations}

Consider a non-LRS orbit in $\mathcal{B}_{\hN_\alpha \hN_\beta}^{\,\mathrm{vac.}}$,
i.e., assume $0 <\vartheta < \pi$. The function $\zeta/\sin\vartheta$
appearing in~\eqref{psieq} is a monotone function,
\begin{equation}\label{zetaovervartheta}
\frac{d}{d\tau_-} \,\frac{\zeta}{\sin\vartheta} =
-\frac{\zeta}{\sin\vartheta} \,(1 + \cos\psi) (1 - \cos\vartheta)\:;
\end{equation}
in particular, for sufficiently small $\tau_-$, $\zeta/\sin\vartheta$ is
bounded from below by a positive constant. The function $\vartheta$
is monotonically decreasing in the reversed direction of time; most importantly,
$\vartheta \rightarrow 0$ as $\tau_- \rightarrow -\infty$.
(Proof: Assume the contrary, i.e., $\vartheta \rightarrow \vartheta_{\infty} =
\mathrm{const}>0$ as $\tau_-\rightarrow -\infty$.%
\footnote{This assumption does not suffice to conclude that $d\vartheta/d\tau_- \rightarrow 0$
  as $\tau_-\rightarrow -\infty$, since the second derivatives are in general not bounded,
  nevertheless, the heuristic reasoning is correct: $d\vartheta/d\tau_-$ approaches zero
  as $\tau_-\rightarrow -\infty$ and thus $\psi$ approaches
  a multiple of $2 \pi$ in this limit, cf.~\eqref{varthetaeq},
  which is a contradiction to~\eqref{psieq}.}
The expression
\begin{equation}\label{twologs}
\frac{d}{d\tau_-} \Big( \log \frac{\zeta}{\sin\vartheta} - \log \tan \frac{\vartheta}{2} \Big)
= -2 ( 1 -\cos \vartheta)
\end{equation}
converges to the limit $c = -2(1 -\cos\vartheta_\infty$) as
$\tau_-\rightarrow -\infty$. We thus obtain the asymptotic
behavior $\zeta \sim e^{- c \tau_-}$ as $\tau_- \rightarrow
-\infty$; accordingly, from~\eqref{psieq}, $d\psi/d\tau_- \sim
-e^{- c \tau_-}$ and hence $\psi \sim e^{-c \tau_-}$ as $\tau_-
\rightarrow -\infty$. Therefore, the integral of $\cos\psi$
behaves asymptotically like the cosine integral
$\mathrm{Ci}(e^{-c \tau_-})$; in particular, the limit exists
as $\tau_-\rightarrow -\infty$. This, however, contradicts the
assumption $\vartheta\rightarrow \vartheta_\infty > 0$ because
of~\eqref{varthetaeq}.) Likewise, for sufficiently small
$\tau_-$, the function $\psi$ is strictly monotonically
decreasing, i.e., $d \psi/d\tau_- < 0$. This is a consequence
of the property $\vartheta\rightarrow 0$ as $\tau_-\rightarrow
-\infty$ (and $\zeta/\sin\vartheta$ being bounded away from
zero).

In the limit $\vartheta \rightarrow 0$, the system~\eqref{polsys}
takes the form
%
\begin{subequations}\label{polsysasy}
\begin{align}
\label{varthetaeqasy}
\frac{d\vartheta}{d\tau_-} & = \textfrac{1}{2} (1-\cos\psi) \vartheta^3 \big(1 + O(\vartheta^2)\big) \:,
& &
\frac{d\zeta}{d\tau_-} & = - \zeta \big[ \cos \psi \; \vartheta^2 + O(\vartheta^4) \big]\:,
\\
\label{psieqasy}
\frac{d\psi}{d\tau_-} & = -4 \vartheta
\sqrt{3 \Big(\frac{\zeta^2}{\sin^2\vartheta} + \frac{1- \cos\psi}{2}\Big)}
+ O(\vartheta^2) \:.
\end{align}
\end{subequations}
We introduce an alternative time variable, $\sigma_-$, by
requiring $d\sigma_- = \vartheta^2 d\tau_-$. Evidently,
$\sigma_-$ is a monotone function of $\tau_-$; most
importantly, $\sigma_- \rightarrow -\infty$ as $\tau_-
\rightarrow -\infty$. (Proof: Assume the contrary, i.e.,
$\sigma_- \rightarrow \sigma_\infty > -\infty$ as
$\tau_-\rightarrow -\infty$. Expressed in $\sigma_-$,
Eq.~\eqref{varthetaeqasy} reads $d\log\vartheta/d\sigma_- =
(1-\cos\psi) \big(1 + O(\vartheta^2)\big)$. However, this
contradicts the fact that $\vartheta\rightarrow 0$ as $\sigma_-
\rightarrow \sigma_\infty$.) Using the variable $\sigma_-$
(which can be identified as a logarithmic time variable), the
system~\eqref{polsysasy} becomes
\begin{subequations}\label{insigma}
\begin{align}
\label{varthetainsigma}
& \frac{d\log\vartheta}{d\sigma_-} = \textfrac{1}{2} (1-\cos\psi) \big(1 + O(\vartheta^2)\big) \,,
& &
\frac{d\log\zeta}{d\sigma_-}  = -\cos \psi + O(\vartheta^2)\:, \\
\label{psiinsigma}
& \frac{d\psi}{d\sigma_-}  = -4 \vartheta^{-1}
\sqrt{3 \Big(\frac{\zeta^2}{\sin^2\vartheta} + \frac{1- \cos\psi}{2}\Big)} + O(\vartheta)\:;
\end{align}
\end{subequations}
in addition,
\begin{equation}
\frac{d}{d\sigma_-} \,\Big( \log \frac{\zeta}{\sin\vartheta}\Big) =
-\frac{1}{2} (1 + \cos\psi) \big( 1 + O(\vartheta^2) \big)\:.
\end{equation}
The function $\zeta/\sin\vartheta$ is monotone,
cf.~\eqref{zetaovervartheta}; furthermore, $\zeta/\sin\vartheta
\rightarrow \infty$ as $\sigma_- \rightarrow -\infty$. (Proof:
Assume the contrary, i.e., $\zeta/\sin\vartheta \rightarrow
a_\infty =\mathrm{const} > 0$ as $\sigma_- \rightarrow
-\infty$. Expressing~\eqref{twologs} in terms of $\sigma_-$ and
integrating this equation yields $\vartheta \sim e^{\sigma_-}$
as $\sigma_- \rightarrow -\infty$. Consistency with the
differential equation for~$\vartheta$,
cf.~\eqref{varthetainsigma}, requires the integral of
$\cos\psi$ as $\sigma_-\rightarrow -\infty$ to converge to
$-1$. This, however, contradicts the differential equation
for~$\psi$, cf.~\eqref{psiinsigma}.)

Finally, we obtain
\begin{equation}
\frac{d\psi}{d\sigma_-} = -4 \sqrt{3}
\,\vartheta^{-1} \frac{\zeta}{\sin\vartheta} \big( 1 + o(1) \big)\:,
\end{equation}
where the function on the r.h.s.\ goes to $-\infty$
monotonically as $\sigma_- \rightarrow -\infty$
(since $\vartheta^{-1}$ and $\zeta/\sin\vartheta$ are monotone and grow
beyond all bounds as $\sigma_- \rightarrow -\infty$).
We conclude that the integral
\begin{equation}
\int_{-\infty}^{\sigma_-} d\sigma_-^\prime \, \cos\psi(\sigma_-^\prime)
\end{equation}
exists. (A statement like this can be regarded as a continuous version
of the Leibniz criterion for alternating series.)
Therefore, by integrating the differential equation for $\zeta$
it follows that $\zeta$ converges
to a positive constant as $\sigma_- \rightarrow -\infty$.
(The solution of the equation for $\vartheta$ implies that
the lower order terms cannot contribute.)

We therefore obtain that the Hubble-normalized variables
$N_\alpha$ and $N_\beta$ converge to (one and the same)
constant as $\tau_-\rightarrow -\infty$; consequently, the
$\alpha$-limit set of a non-LRS orbit is a point with
$\Sigma_\alpha = \Sigma_\beta = -1$, $\Sigma_\gamma =2$,
$N_\alpha = N_\beta > 0$ (and $N_\gamma = 0$). Expressed in
terms of the dominant variables, the $\alpha$-limit set of a
non-LRS orbit in $\mathcal{B}_{\hN_\alpha
\hN_\beta}^{\,\mathrm{vac.}}$ is a point on
$\mathrm{TL}_\gamma$.
\end{proof}

\begin{remark}
Based on this information it is straightforward to consider and
analyze the asymptotic system of differential equations, which
arises by inserting the asymptotic behavior of solutions
into~\eqref{polsys}. The asymptotic oscillations of solutions
can be read off directly.
\end{remark}

\textbf{The fluid case} $\bm{\hOmega > 0}$. On the representative
$\mathcal{B}_{\hN_\alpha \hN_\beta}^{\,\mathrm{fl.}}$ of $\BVII^{\mathrm{fl.}}$
we adapt a monotone function found
by Uggla, given in~\cite{waiell97},
\begin{equation}\label{monfunzeta0}
\zeta_0 = \left( 2 \hH - v \hSigma_\gamma \right)^{-2(1+v)}
\hOmega \:(\hN_\alpha \hN_\beta)^v  \qquad
\text{with} \text \quad v = \sfrac{1}{4} (1 +3 w)\:.
\end{equation}
Note that $0< v <1$ since $-\sfrac{1}{3} < w < 1$; hence $2 \hH
- v \hSigma_\gamma >0$ if $\hH > 0$ (since $\hSigma_\gamma \in
(-2 \hH, 2 \hH)$ because of the constraints) and thus $0\leq
\zeta_0 < \infty$ on
$\overline{\mathcal{B}}_{\hN_\alpha\hN_\beta} \backslash
\{\mathrm{Z}_\gamma\}$. On this subset we find
\begin{align*}
\frac{d}{d\htau} \zeta_0 & =  -\left( 2 \hH - v \hSigma_\gamma \right)^{-1}
\zeta_0 \left[ \sfrac{2}{3} (1 - v^2) (\hSigma_\alpha- \hSigma_\beta)^2 +
2 \left( 2 v \hH - \hSigma_\gamma \right)^2 \right], \\
\frac{d^3\zeta_0}{d\htau^3} \Big|_{d\zeta_0/d\htau = 0} & =
-\left( 2 \hH - v \hSigma_\gamma \right)^{-1} \zeta_0
\left[ \sfrac{4}{3} (1-v^2) (\hN_\alpha^2 - \hN_\beta^2)^2 +
(1-v)^2 \left( 8 v \hOmega -\sfrac{2}{3}
(\hN_\alpha-\hN_\beta)^2 \right)^2 \right],
\end{align*}
hence $\zeta_0$ is monotonically decreasing on
$\mathcal{B}_{\hN_\alpha\hN_\beta}^{\,\mathrm{fl.}}$.

This allows us to apply the monotonicity principle:
Using that $\zeta_0 = 0$ on $\overlineBII$ and on the vacuum
subset $\mathcal{B}_{\hN_\alpha\hN_\beta}^{\,\mathrm{vac.}}$
we conclude that the $\alpha$-limit of every orbit in
$\mathcal{B}_{\hN_\alpha\hN_\beta}^{\,\mathrm{fl.}}$ is one of
the fixed points $\mathrm{Z}_\gamma$.
The asymptotic approach to the fixed points $\mathrm{Z}_\gamma$
provides an example for asymptotic self-similarity breaking,
see~\cite{waietal99} for details and explicit decay rates.
(Note that the points $\mathrm{Z}_\gamma$ do not appear 
as fixed points in the Hubble-normalized approach; instead, 
they are associated with `infinity' 
in the Hubble-normalized state space and thus
do not correspond to self-similar solutions).

The results on the $\omega$-limit sets of type~$\mathrm{VII}_0$
orbits are not needed in the proof of the Bianchi type~IX
attractor theorem below. However, for completeness we now
analyze the possible $\omega$-limit sets of orbits in
$\mathcal{B}_{\hN_\alpha\hN_\beta}^{\,\mathrm{fl}}$. The
monotonicity principle, applied to the
function~\eqref{monfunzeta0}, implies that the $\omega$-limit
sets of orbits must be contained in
$\overline{\mathcal{B}}_{\hN_\alpha}$ or
$\overline{\mathcal{B}}_{\hN_\beta}$ (which are part of
$\overlineBII$) or/and in
the vacuum subset $\mathcal{B}_{\hN_\alpha\hN_\beta}^{\,\mathrm{vac.}}$. 
However, on
$\mathcal{B}_{\hN_\alpha\hN_\beta}^{\,\mathrm{vac.}}$ only
$\mathrm{TL}_\alpha$ is admissible. To prove this assume that
there exists an orbit $\gamma$ that has an $\omega$-limit point
$\mathrm{P}$ on
$\mathcal{B}_{\hN_\alpha\hN_\beta}^{\,\mathrm{vac.}}$ 
and that this point does not lie on $\mathrm{TL}_\gamma$. Then
the orbit through $\mathrm{P}$ and the $\omega$-limit set of
that orbit must be contained in $\omega(\gamma)$. As proved
above, this $\omega$-limit set consists of one single fixed
point $\mathrm{K}_{\mathrm{P}}$ on the Kasner circle (which
acts as a sink, when viewed as a fixed point on the [closure of
the] vacuum subset). Since $\hOmega^{-1}d\hOmega/d\htau =
-3(1-w)$ on the Kasner circle, $\mathrm{K}_{\mathrm{P}}$ is a
(transversally hyperbolic) sink also when viewed as a fixed
point on $\overline{\mathcal{B}}_{\hN_\alpha\hN_\beta}$. The
conclusion that the sink $\mathrm{K}_{\mathrm{P}}$ is in
$\omega(\gamma)$ contradicts the assumption
$\mathrm{P}\in\omega(\gamma)$. This leaves only the fixed
points on $\mathrm{TL}_\gamma$ as possible $\omega$-limits on
the vacuum subset
$\mathcal{B}_{\hN_\alpha\hN_\beta}^{\,\mathrm{vac.}}$. In fact,
for each fixed point $\mathrm{L}_\gamma \in \mathrm{TL}_\gamma$
there exists exactly one orbit in $\mathcal{B}_{\hN_\alpha\hN_\beta}^{\,\mathrm{fl.}}$ 
converging to $\mathrm{L}_\gamma$ as $\htau \rightarrow
\infty$. This follows from the center manifold theorem by
noting that the fixed point $\mathrm{L}_\gamma$ acts as a
center saddle (which is because $\hOmega^{-1} d\hOmega/d\htau =
-3(1-w) \hH$ along $\mathrm{TL}_\gamma$). One easily identifies
the orbit that converges to $\mathrm{L}_\gamma$ as being an LRS
orbit (the LRS case is exactly solvable).
%
%
We have thus shown that convergence to $\mathrm{TL}_\gamma$
occurs for a non-generic set of orbits; the $\omega$-limit set
of every other orbit lies on the sets
$\overline{\mathcal{B}}_{\hN_\alpha}$ and
$\overline{\mathcal{B}}_{\hN_\beta}$ (and is thus of type~I
or~II).

The details are as follows: The remaining LRS solutions either
converge to $\mathrm{Q}_\gamma$ (a one-parameter family) or
converge to $\mathrm{F}$ as $\htau\rightarrow \infty$ (one
solution). The local analysis of these fixed points, see
Section~\ref{nongeneric} for the general case, implies that the
former are embedded into a two-parameter families of orbits
converging to $\mathrm{Q}_\gamma$, while the latter are
embedded into a one-parameter families of orbits converging to
$\mathrm{F}$. Furthermore, there exist two (equivalent) orbits
converging to each of the fixed points $\mathrm{CS}_\alpha$,
$\mathrm{CS}_\beta$.\footnote{Equivalence of orbits refers to
the discrete symmetries of the problem that are associated with
interchanging the axes.} The generic scenario, however, is
convergence to one of the transversally hyperbolic sinks on the
Kasner circle. To prove this statement suppose that there
exists an orbit $\gamma$ that possesses an $\omega$-limit point
$\mathrm{P}$ on $\mathcal{B}_{\hN_\alpha}$
($\mathcal{B}_{\hN_\beta}$) with $\mathrm{P} \neq
\mathrm{CS}_\alpha$ ($\mathrm{P}\neq \mathrm{CS}_\beta$). Then
the orbit through $\mathrm{P}$ and the $\alpha$-limit set of
that orbit must be contained in $\omega(\gamma)$. As stated in
Section~\ref{typeIIsubset}, this $\alpha$-limit set coincides with 
$\mathrm{CS}_\alpha$ ($\mathrm{CS}_\beta$) in
$\mathcal{B}_{\hN_\alpha}$ ($\mathcal{B}_{\hN_\beta}$).
As this point 
is a hyperbolic saddle in
$\overline{\mathcal{B}}_{\hN_\alpha\hN_\beta}$, the orbit
converging to $\mathrm{CS}_\alpha$ ($\mathrm{CS}_\beta$) as
$\htau\rightarrow \infty$ must be contained in $\omega(\gamma)$
as well. However, this is a contradiction to the fact that
$\omega(\gamma)$ is a subset of
$\overline{\mathcal{B}}_{\hN_\alpha} \cup
\overline{\mathcal{B}}_{\hN_\beta}$. Analogously, one can prove
that $\omega(\gamma)$ and the interior of $\BI^{\mathrm{fl.}}$
are disjoint, which leads to the statement. Summing up, the
$\omega$-limit of a generic orbit in $\BVII^{\mathrm{fl.}}$ is
one of the transversally hyperbolic sinks on the Kasner circle.

\section{Non-generic solutions: Asymptotic self-similarity}
\label{nongeneric}

In the previous section we have identified the fixed points
associated with the system~\eqref{domsys} on the Bianchi
boundary subsets of $\overlineBIX$. A local dynamical systems
analysis of the fixed points shows whether or not these points
attract type~IX orbits in the limit $\htau\rightarrow \infty$.
\begin{itemize}
\item[$\mathrm{K}^{\ocircle}$]
Evaluated on the Kasner circle, Eq.~\eqref{domsys} implies
$\hN_\alpha^{-1}d\hN_\alpha/d\htau\,|_{\mathrm{K}^{\ocircle}} = -2( 1+
\hSigma_\alpha)$ ($\alpha=1,2,3$) and
$\hOmega^{-1}d\hOmega/d\htau\,|_{\mathrm{K}^{\ocircle}} = -3(1-w)$. Each fixed
point $\mathrm{K}$ on $\mathrm{K}^{\ocircle}
\backslash\{\mathrm{T}_1,\mathrm{T}_2, \mathrm{T}_3\}$ is a
transversally hyperbolic saddle
that has one unstable mode
and three stable modes. The unstable manifold of
$\mathrm{K}$ coincides with a vacuum type~II orbit, see
Figure~\ref{alltypeII}; the three-dimensional stable
manifold is contained in $\overlineBVII$. (The
one-dimensional center manifold is $\mathrm{K}^\ocircle$
itself.) Therefore, there do not exist any type~IX
solutions that converge to $\mathrm{K}$ as
$\htau\rightarrow \infty$. The \textit{Taub points}
$\{\mathrm{T}_1,\mathrm{T}_2, \mathrm{T}_3\}$ are not
transversally hyperbolic. Each Taub point
$\mathrm{T}_\alpha$ possesses a two-dimensional stable
manifold and a three-dimensional center manifold. The
(closure of the) two-dimensional stable manifold of
$\mathrm{T}_\alpha$ coincides with the LRS subset of
$\overline{\mathcal{B}}_{\hN_\alpha}$ (which contains the
two special orbits $\mathrm{Q}_\alpha \rightarrow
\mathrm{T}_\alpha$ and $\mathrm{F} \rightarrow
\mathrm{T}_\alpha$ on the vacuum subset of
$\mathcal{B}_{\hN_\alpha}$
and on the Bianchi type~I fluid subset, 
respectively.) The three-dimensional center manifold
coincides with the vacuum subset of
$\overline{\mathcal{B}}_{\hN_\beta \hN_\gamma}$ (which is a
vacuum $\overlineBVII$ set). Therefore, the center manifold
reduction theorem~\cite{cra91} reduces the problem to
analyzing Bianchi type~$\mathrm{VII}_0$ vacuum dynamics. In
Section~\ref{Bianchisevenzero} we have shown that
$\mathrm{T}_\alpha$ is excluded as an $\omega$-limit point
for orbits in $\overline{\mathcal{B}}_{\hN_\beta
\hN_\gamma}^{\,\mathrm{vac.}}$; the existence of the
monotone function~\eqref{monvii0} implies that
$\mathrm{T}_\alpha$ is a center saddle in $\overlineBIX$.
Consequently, there do not exist any type~IX solutions that
converge to any of the Taub points as $\htau\rightarrow
\infty$.
\item[F\,\,] Eq.~\eqref{IXeq} implies that
    $\hSigma_\alpha^{-1}d
    \hSigma_{\alpha}/d\htau\,|_{\mathrm{F}} = \sfrac{3}{2}
    (1-w)$ and $\hN_\alpha^{-1} d
    \hN_{\alpha}/d\htau\,|_{\mathrm{F}} = -\sfrac{1}{2} (1
    + 3 w)$. Therefore, $\mathrm{F}$ is a hyperbolic saddle
    that possesses a two-dimensional unstable manifold,
    which coincides with the
Bianchi type~I subset, 
and a three-dimensional stable manifold.
Accordingly, $\mathrm{F}$
attracts a two-parametric family of type~IX orbits as
$\htau\rightarrow \infty$. These solutions have a so-called
isotropic singularity.
\item[$\mathrm{CS}_\alpha$]
The fixed points $\mathrm{CS}_\alpha$ ($\alpha = 1,2,3$) are
hyperbolic with a three-dimensional unstable and a two-dimensional
stable manifold; the former coincides with
$\mathcal{B}_{\hN_\alpha}$, the latter is associated with the equations
$\hN_\beta^{-1} d\hN_\beta/d\htau\,|_{\mathrm{CS}_\alpha} = \sfrac{3}{4} (1+3 w)$
(for $\beta \neq \alpha$). Therefore, each of the fixed points
$\mathrm{CS}_\alpha$ attracts an (equivalent) one-parameter set of
type~IX orbits in the limit $\htau\rightarrow \infty$.
\item[$\mathrm{TL}_\alpha$] Each fixed point of the line
    $\mathrm{TL}_\alpha$ has a three-dimensional center
    manifold and a two-di\-men\-sional stable manifold. The
    center manifold coincides with the vacuum subset
    $\overline{\mathcal{B}}_{\hN_\beta
    \hN_\gamma}^{\,\mathrm{vac.}}$; in
    Section~\ref{Bianchisevenzero} we have proved that the
    points of $\mathrm{TL}_\alpha$ take the role of
    sources. We thus conclude that the fixed points on
    $\mathrm{TL}_\alpha$ are center saddles. The
    two-dimensional stable manifold of each point of
    $\mathrm{TL}_\alpha$ is contained in the LRS subset
    $\mathcal{L\!R\!S}_\alpha$ of $\BIX$ (which is the
    hyperplane given by the conditions $\hSigma_\beta =
    \hSigma_\gamma$ and $\hN_\beta = \hN_\gamma$); more
    specifically, the closure of the union of the unstable
    manifolds coincides with the closure of
    $\mathcal{L\!R\!S}_\alpha$. Therefore, for each fixed
    point on $\mathrm{TL}_\alpha$ there exists a
    one-parameter family of type~IX orbits that converges
    to this point as $\htau\rightarrow \infty$; these
    orbits correspond to LRS solutions. (Conversely,
    generic LRS type~IX solutions converge to
    $\mathrm{TL}_\alpha$, see e.g. \cite{waiell97}.)
\end{itemize}

The solutions whose $\omega$-limit is one of the fixed points
form a subfamily of measure zero of the (four-parameter) family
of Bianchi type~IX solutions. Following the nomenclature
of~\cite{rin01} we thus refer to these solutions as
\textit{non-generic} solutions of Bianchi type~IX.
Alternatively, to capture the asymptotic behavior of these
solution, we use the term \textit{past asymptotically
self-similar} solutions. (Since a fixed point in the
Hubble-normalized dynamical systems formulation corresponds to
a self-similar solution, see e.g.~\cite{waiell97}, solutions
that converge to one of the above fixed points are
asymptotically self-similar.)

Apart from the invariant Bianchi contraction subsets there exists
other invariant subsets of the full state space. 
The most important are the three equivalent LRS subsets
$\mathcal{L\!R\!S}_\gamma$ defined by $\hSigma_\alpha
=\hSigma_\beta$ and $\hN_\alpha= \hN_\beta$ for
$(\alpha\beta\gamma) \in \{(123),(231),(312)\}$. The past
asymptotically self-similar solutions comprise the LRS Bianchi
type~IX solutions. As seen above, generic LRS solutions
converge to $\mathrm{TL}_\alpha$ towards the past (and each
solution that converges to $\mathrm{TL}_\alpha$ is LRS), but
there exist exceptional LRS solutions that converge to
$\mathrm{F}$ or $\mathrm{CS}_\alpha$. The remaining orbits
whose limit point is either $\mathrm{F}$ or
$\mathrm{CS}_\alpha$ correspond to past asymptotically
self-similar solutions that are non-LRS.
Clearly, every solution that converges to $\mathrm{F}$ or
$\mathrm{CS}_\alpha$ is a non-vacuum solution, since $\hOmega \neq 0$
at $\mathrm{F}$ and $\mathrm{CS}_\alpha$.

It is natural to ask how the non-generic orbits are embedded in
the state space $\overlineBIX$. The LRS orbits form the three
LRS subsets $\mathcal{L\!R\!S}_\alpha$, which are the
hyperplanes given by the conditions $\hSigma_\beta =
\hSigma_\gamma$, $\hN_\beta = \hN_\gamma$, where
$(\alpha\beta\gamma) \in \{(123),(231),(312)\}$. The orbits
whose $\omega$-limit set is the fixed point
$\mathrm{CS}_\alpha$ (for some $\alpha$) form the set
$\mathcal{C\!S}_\alpha$ in $\BIX$; we call
$\mathcal{C\!S}_\alpha$ the Collins-Stewart manifold. The local
analysis of the fixed point $\mathrm{CS}_\alpha$ and the
regularity of the dynamical system~\eqref{IXeq} imply that the
Collins-Stewart manifold $\mathcal{C\!S}_\alpha$ is a
two-dimensional surface; it can be viewed as a two-dimensional
manifold with boundary embedded in $\overlineBIX$ (where this
boundary lies in $\overlineBVII$). Analogously, the orbits
whose $\alpha$-limit set is the fixed point $\mathrm{F}$ form
the set $\mathscr{F}$ in $\BIX$, which we call the isotropic
singularity manifold, since solutions converging to
$\mathrm{F}$ are those with an isotropic singularity. The
isotropic singularity manifold $\mathscr{F}$ is a
three-dimensional hypersurface; it can be viewed as a
three-dimensional manifold with boundary.

In the subsequent section we will state and prove
the Bianchi type~IX attractor theorem, which concerns
the behavior of generic Bianchi type~IX models
(i.e., those that are not asymptotic
self-similar, which provides an example of asymptotic self-similarity
breaking; for other such examples, see~\cite{limetal06}).

\section{A new proof of the Bianchi type IX attractor theorem}
\label{newproof}

\begin{definition}
Consider a solution of Bianchi type~IX that is either vacuum or
associated with a perfect fluid satisfying $-\textfrac{1}{3}< w < 1$. Such a
solution is called\/ \emph{generic} if it is not past
asymptotically self-similar.
\end{definition}

\begin{remark}
Accordingly, a solution is generic if its $\omega$-limit set is
neither the point $\mathrm{F}$, nor any of the points
$\mathrm{CS}_\alpha$, nor a point on $\mathrm{TL}_\alpha$;
in other words, a generic
solution corresponds to an orbit in $\BIX$ that is neither contained
in the submanifold $\mathscr{F}$, nor in $\mathcal{C\!S}_\alpha$, nor in
the hyperplane $\mathcal{L\!R\!S}_\alpha$. Therefore, the set of generic Bianchi type~IX
states is an open set in $\BIX$.
\end{remark}
%

\begin{definition}
The\/ \emph{Mixmaster attractor} $\mathcal{A}_{\mathrm{IX}}$
(alternatively referred to as the Bianchi type~IX attractor) is
defined to be the subset of $\overlineBIX$ given by
union of the Bianchi type~I and~II vacuum subsets, i.e.,
$\mathcal{A}_{\mathrm{IX}} = \BI^{\mathrm{vac.}} \cup \BII^{\mathrm{vac.}}$.
Accordingly, $\mathcal{A}_{\mathrm{IX}}$ consists of the three
representations of the Bianchi type~II vacuum subset and the Kasner
circle (the Bianchi type~I vacuum subset), i.e.,
\begin{equation}\label{AIXdef}
\mathcal{A}_{\mathrm{IX}} = \mathrm{K}^\ocircle \cup
\mathcal{B}_{N_1}^{\mathrm{vac.}} \cup
\mathcal{B}_{N_2}^{\mathrm{vac.}} \cup
\mathcal{B}_{N_3}^{\mathrm{vac.}} \:.
\end{equation}
\end{definition}

The main result concerning generic Bianchi type~IX models is the Bianchi
type~IX attractor theorem, which is due
to Ringstr\"om~\cite{rin01}; this result rests on earlier work that
is reviewed and derived in~\cite{waiell97}, and on~\cite{ren97}
and~\cite{rin00}. In the following we state the Bianchi type~IX
attractor theorem in a version adapted to our purposes.

\begin{theorem}[\cite{rin01}]\label{rinthm}
A generic orbit $\gamma$ in $\BIX$ has an $\omega$-limit set that is
a subset of the Mixmaster attractor, i.e.,
$\omega(\gamma) \subseteq \mathcal{A}_{\mathrm{IX}} = \BI^{\mathrm{vac.}} \cup \BII^{\mathrm{vac.}}$.
\end{theorem}

\begin{remark}
Note that we have chosen the time direction towards the past singularity,
see~\eqref{newtime} and~\eqref{domtime}.
When we use the standard future directed time, `$\omega$-limit set'
is replaced by `$\alpha$-limit set'.
\end{remark}

\begin{remark}
For an equivalent formulation of Theorem~\ref{rinthm}
let $X(\htau) =
(\hSigma_1,\hSigma_2,\hSigma_3,\hN_1,\hN_2,\hN_3,\hH)(\htau)$ be a generic
solution of Bianchi type~IX. Then
\begin{equation}\label{distfromatt}
\| X(\htau) - \mathcal{A}_{\mathrm{IX}} \| \rightarrow 0 \qquad
(\htau\rightarrow \infty)\:,
\end{equation}
where the distance $\|X -  \mathcal{A}_{\mathrm{IX}}\|$ is given as
$\min_{Y\in\mathcal{A}_{\mathrm{IX}}} \| X- Y\|$.
\end{remark}

\begin{remark}
Alternatively, the statement of the theorem can be expressed as follows:
Along every generic orbit we have
\begin{equation}\label{prevform}
\hN_1 \hN_2 + \hN_1 \hN_3 + \hN_2 \hN_3 \rightarrow 0
\quad\text{and}\quad
\hOmega \rightarrow 0
\quad \text{as } \htau \rightarrow \infty\:.
\end{equation}
In combination with the constraint~\eqref{Hconstraint} we further obtain
$\bar{H}\rightarrow 1$ as $\htau \rightarrow \infty$. Since, by definition,
$\bar{H}= \left[1 + \textfrac{1}{6}\big( N_1 N_2 + N_1 N_3 + N_2 N_3 \right]^{-1/2}$,
see~\eqref{gauss} and~\eqref{Ddef},
it follows that $\big( N_1 N_2 + N_1 N_3 + N_2 N_3\big) \rightarrow 0$
when $\htau \rightarrow \infty$.
An immediate consequence, cf.~\eqref{domvars}, is
that $\Omega/\hOmega\rightarrow 1$ and $d\tau_-/d\htau \rightarrow 1$ as
$\htau \rightarrow \infty$, so that $\tau_- \rightarrow \infty$
as $\htau \rightarrow \infty$. Using these results, Theorem~\ref{rinthm}
instantly yields the original formulation of this theorem in~\cite{rin01}.
\end{remark}

In the following we give a new proof of Ringstr\"om's Bianchi
type~IX attractor theorem.
The proof we present is subdivided into a number of lemmas,
which culminate in Theorem~\ref{rinthm}.
The additional remarks in this subsection are not needed directly
for the proofs, but give further insights about the
main ideas and the lines of argument.

\begin{lemma}\label{monlemma}
Every orbit in $\BIX$ possesses a non-empty $\omega$-limit set that
is contained in the subset $\overlineBVII$ of the boundary
$\partial\BIX$.
\end{lemma}
%


%
\begin{proof}
Since the state space $\BIX$ is relatively compact, every orbit has
an $\omega$-limit point in $\overlineBIX = \overline{\bm{H}}_0 \cup \overlineBVII$, cf.~\eqref{partialBnine}.
The function%
%
%
%
\begin{equation}
\bar{\Delta} := \hH^{-3} \hN_1 \hN_2 \hN_3\:
\end{equation}
is positive and strictly monotonically decreasing on $\BIX$, since
\begin{equation}\label{Mdecr}
\frac{d\bar{\Delta}}{d\htau}  = - 3 \:\frac{\hq}{\hH}\:
\bar{\Delta}\:,
\quad
\frac{d^2\bar{\Delta}}{d\htau^2}\, \Big|_{\hSigma^2 = 0} = 0\:,
\quad
\frac{d^3\bar{\Delta}}{d\htau^3}\, \Big|_{\hSigma^2 = 0} = -
2 \hH^{-1}\left( {}^3\!\hS_1^2 + {}^3\!\hS_2^2 + {}^3\!\hS_3^2
\right)\bar{\Delta}\:;
\end{equation}
in the perfect fluid case the first derivative is negative
since $\hq = 2\hSigma^2 + \frac{1}{2}(1+3w) \hOmega>0$; in the
vacuum case, i.e., $\hOmega =0$, $d \bar{\Delta}/d \htau = 0$
is possible since $\hq = 0$ when $\hSigma^2 = 0$, but then the
third derivative is negative, since ${}^3\!\hS_1^2 +
{}^3\!\hS_2^2 + {}^3\!\hS_3^2 > 0$ when $\hSigma^2 = 0$ because
of the constraints. Application of the monotonicity principle
to the function $\bar{\Delta}$ yields that the $\omega$-limit
set of every orbit is contained in $\partial \BIX$, where the
set $\bm{H}_0$ is excluded because $\bar{\Delta} = \infty$ on
$\bm{H}_0$.
\end{proof}

The set $\overlineBVII$ can be decomposed according to
\begin{equation}\label{BVIIsplit}
\overlineBVII = \BVII \cup \BII \cup \BI \cup \{\mathrm{Z}_1,\mathrm{Z}_2,\mathrm{Z}_3\}\:.
\end{equation}
In the subsequent lemmas we exclude the possibility that
generic orbits have $\omega$-limits in $\BVII$ or
$\{\mathrm{Z}_\alpha\}$; for pedagogical reasons some of the
lemmas refer to the (simpler) vacuum case and the fluid case
separately. Furthermore, we prove that not only in the vacuum
case but also in the fluid case only the vacuum subsets of
$\BI$ and $\BII$ come into question.
Taken together these lemmas then directly lead to 
Theorem~\ref{rinthm}.

\begin{lemma}[Vacuum case]\label{noVII}
A generic orbit in $\BIX$ cannot have an $\omega$-limit point on the
subset $\BVII$.
\end{lemma}
\begin{proof}(\textit{Vacuum case, i.e., $\hOmega =0$}. In this case
`generic' reduces to non-LRS, cf.~Section~\ref{nongeneric}.)
In preparation for the proof we begin by considering
a fixed point $\mathrm{L}_\alpha$ of the line
$\mathrm{TL}_\alpha$, see~\eqref{TLdef};
in particular, $\hN_\alpha = 0$, and
$\hN_\beta = \hN_\gamma> 0$ at $\mathrm{L}_\alpha$.
The point $\mathrm{L}_\alpha$ has the following properties:
$\mathrm{L}_\alpha$ is not
hyperbolic, but on its center manifold, which is the set
$\BVII$, there exists a one-parameter
set of orbits converging to $\mathrm{L}_\alpha$ as $\htau
\rightarrow -\infty$; in contrast, no orbit converges to
$\mathrm{L}_\alpha$ as $\htau \rightarrow \infty$;
see the previous discussion of Bianchi type~$\mathrm{VII}_0$ models.
In this sense, $\mathrm{L}_\alpha$ resembles a (transversal) source for
$\BVII$. Furthermore, from~\eqref{Nalpha}
we obtain that $\hN_\alpha^{-1} d \hN_\alpha/d\htau = -6 \hH$
at $\mathrm{L}_\alpha$,
%
%
hence $\mathrm{L}_\alpha$ possesses a one-dimensional stable
subspace that lies in $\BIX$.
The center manifold reduction theorem applies~\cite{cra91}:
There exists a neighborhood of $\mathrm{L}_\alpha$ such that the flow
of the full nonlinear system is equivalent to the flow of the
decoupled system
\begin{subequations}\label{Lalphasaddle}
\begin{align}
\frac{d}{d\htau} \hN_\alpha & = - 6 \hH \hN_\alpha \\
\frac{d}{d\htau} B_{\mathrm{VII}_0} & =
F(B_{\mathrm{VII}_0})\:,
\end{align}
\end{subequations}
where $B_{\mathrm{VII}_0}$ denotes the collection of the
variables of $\BVII$. Since $\mathrm{L}_\alpha$ is a
(transversal) source for the second subsystem,
$\mathrm{L}_\alpha$ is a center saddle and there exists exactly
one orbit whose $\omega$-limit is $\mathrm{L}_\alpha$; this
orbit coincides with the unstable manifold of
$\mathrm{L}_\alpha$ (for which $\hN_\alpha > 0$, $\hN_\beta >
0$, $\hN_\gamma >0$) and can straightforwardly be identified as
a Bianchi type~IX LRS orbit (because the vacuum type IX LRS
subset is exactly solvable).

Now consider a non-LRS orbit $\gamma$ in the vacuum subset of
$\BIX$ and assume that the $\omega$-limit set $\omega(\gamma)$
contains a point on $\BVII$, i.e., a point $\mathrm{P}$ such
that $\hN_\alpha = 0$, $\hN_\beta > 0$, $\hN_\gamma > 0$. We
distinguish three possible cases: (i)
$\mathrm{P}\in\mathrm{TL}_\alpha$; (ii) $\mathrm{P} \in \BVII$,
$\mathrm{P} \not\in \mathrm{TL}_\alpha$,
$\mathrm{P}\not\in\mathcal{QL}_\alpha$; (iii) $\mathrm{P} \in
\mathcal{QL}_\alpha$.

Consider case (i), i.e., assume $\mathrm{P} = \mathrm{L}_\alpha
\in \omega(\gamma)$. There are two possibilities: Either
$\omega(\gamma) = \{ \mathrm{L}_\alpha\}$, then $\gamma$ is the
orbit that coincides with the stable manifold of
$\mathrm{L}_\alpha$; this is impossible since the stable
manifold is an LRS orbit; or $\omega(\gamma) \supsetneq \{
\mathrm{L}_\alpha\}$. The saddle structure of
$\mathrm{L}_\alpha$ allows us to draw the following conclusion:
Since $\{\mathrm{L}_\alpha\} \subsetneq \omega(\gamma)$, it
follows that also the stable manifold of $\mathrm{L}_\alpha$
must be a subset of $\omega(\gamma)$. However, since the stable
manifold lies in $\BIX$, this is a contradiction to
Lemma~\ref{monlemma}. Hence, the existence of a point
$\mathrm{P} \in \mathrm{TL}_\alpha$ in $\omega(\gamma)$ is
excluded for the non-LRS orbit $\gamma$.

Case (ii) is analogous. Since $\mathrm{P}$ is contained in
$\omega(\gamma)$, so is the entire Bianchi type
$\mathrm{VII}_0$ orbit through $\mathrm{P}$ and therefore the
$\alpha$- and $\omega$-limit sets of that orbit. The
$\alpha$-limit set is a fixed point $\mathrm{L}_\alpha$ on the
line $\mathrm{TL}_\alpha$, cf.~the analysis of Bianchi
type~$\mathrm{VII}_0$ models; accordingly, $\mathrm{L}_\alpha
\supsetneq \omega(\gamma)$ which leads to a contradiction to
Lemma~\ref{monlemma} in analogy with case (i).

Finally, consider case (iii). Since $\mathrm{P} \in
\omega(\gamma)$, the entire orbit through $\mathrm{P}$ (and the
$\alpha$- and $\omega$-limit points of that orbit) must be
contained in $\omega(\gamma)$ as well; hence,
$\overline{\mathcal{QL}}_\alpha \subseteq \omega(\gamma)$.
Because we have already excluded cases (i) and (ii),
we know that 
$\omega(\gamma) \subseteq \overlineBII \cup
\big(\bigcup_{\delta} \overline{\mathcal{QL}}_\delta\big)$.
However, the heteroclinic orbits $\mathcal{QL}_\delta$, $\delta=1,2,3$,
are not connected, but `isolated branches'
of the set $\overlineBII \cup
\big(\bigcup_{\delta} \overline{\mathcal{QL}}_\delta\big)$.
Since such structures can never be part
of a limit set, cf.~the remark below,
the assumption $\mathcal{QL}_\alpha \subseteq \omega(\gamma)$ and
therefore (iii) result in a contradiction.
This finishes the proof of the lemma.
\end{proof}

\begin{remark}
Let us elaborate on case (iii) in the proof of the theorem and
show explicitly why the isolated heteroclinic orbits
$\mathcal{QL}_\delta$, $\delta=1,2,3$, are excluded from the
possible $\omega$-limit set. In the proof of Lemma~\ref{noVII},
case~(iii), we assume that there exists an orbit $\gamma$ such
that $\mathrm{P}\in\mathcal{QL}_\alpha$ is an element of
$\omega(\gamma)$; hence there exists a diverging sequence of
times, $(\varsigma_n)_{n\in\mathbb{N}}$, such that
$\gamma(\varsigma_n) \rightarrow \mathrm{P}$ ($n\rightarrow
\infty$). Let $\mathscr{V}$ be a sufficiently small
neighborhood of $\mathrm{Z}_\alpha$, preferably generated by an
open ball, so that $\mathrm{P} \not\in \mathscr{V}$. For
sufficiently large $n$, there exists times $\sigma_n$,
$\varsigma_{n-1} < \sigma_n < \varsigma_n$, such that
$\gamma(\sigma_n) \in \mathscr{V}$. (This is a simple
consequence of the continuous dependence of the flow on initial
data; recall that $\mathcal{QL}_\alpha$ is a heteroclinic orbit
connecting the fixed point $\mathrm{Z}_\alpha$ with the fixed
point $\mathrm{Q}_\alpha$.) Consequently, there exist times
$\kappa_n$, $\varsigma_{n-1} < \kappa_n < \varsigma_n$, such
the orbit $\gamma$ enters $\mathscr{V}$ at $\kappa_n$ (i.e.,
$\gamma(\kappa_n) \in \partial\mathscr{V}$,
$\gamma(\kappa_n+\epsilon) \in \mathscr{V}$ for sufficiently
small $\epsilon > 0$). For all $n$, $\gamma(\kappa_n)$ is
contained in the complement of a sufficiently small
neighborhood $\mathscr{U}$ of the point $\partial\mathscr{V}
\cap \mathcal{QL}_\alpha$ (because the flow of the dynamical
system points out of $\mathscr{V}$ in $\partial\mathscr{V}\cap
\mathscr{U}$). By going over to a subsequence
$\gamma(\kappa_n)$, this implies that $\gamma$ has a
$\omega$-limit point on $\partial\mathscr{V} \backslash
\mathscr{U}$, i.e., an $\omega$-limit point in $\BVII$ (or
$\bm{H}_0$) that is not contained on $\mathcal{QL}_\delta$,
$\delta=1,2,3$. This is a contradiction to the assumption
$\omega(\gamma) \subseteq \overlineBII \cup
\big(\bigcup_{\delta} \overline{\mathcal{QL}}_\delta\big)$.
\end{remark}

\begin{figure}[Ht]
\psfrag{BII}[lc][lc][1.2][0]{$\overlineBII$}
\psfrag{Z1}[cc][cc][1][0]{$\mathrm{Z}_1$}
\psfrag{Z2}[cc][cc]{$\mathrm{Z}_2$}
\psfrag{Z3}[bc][cc]{$\mathrm{Z}_3$}
\psfrag{B1}[rc][rc]{$\mathcal{B}_{\hN_2 \hN_3}$}
\psfrag{B2}[lc][lc]{$\mathcal{B}_{\hN_1 \hN_3}$}
\psfrag{B3}[lc][lc]{$\mathcal{B}_{\hN_1 \hN_2}$}
\centering{
\includegraphics[height=0.35\textwidth]{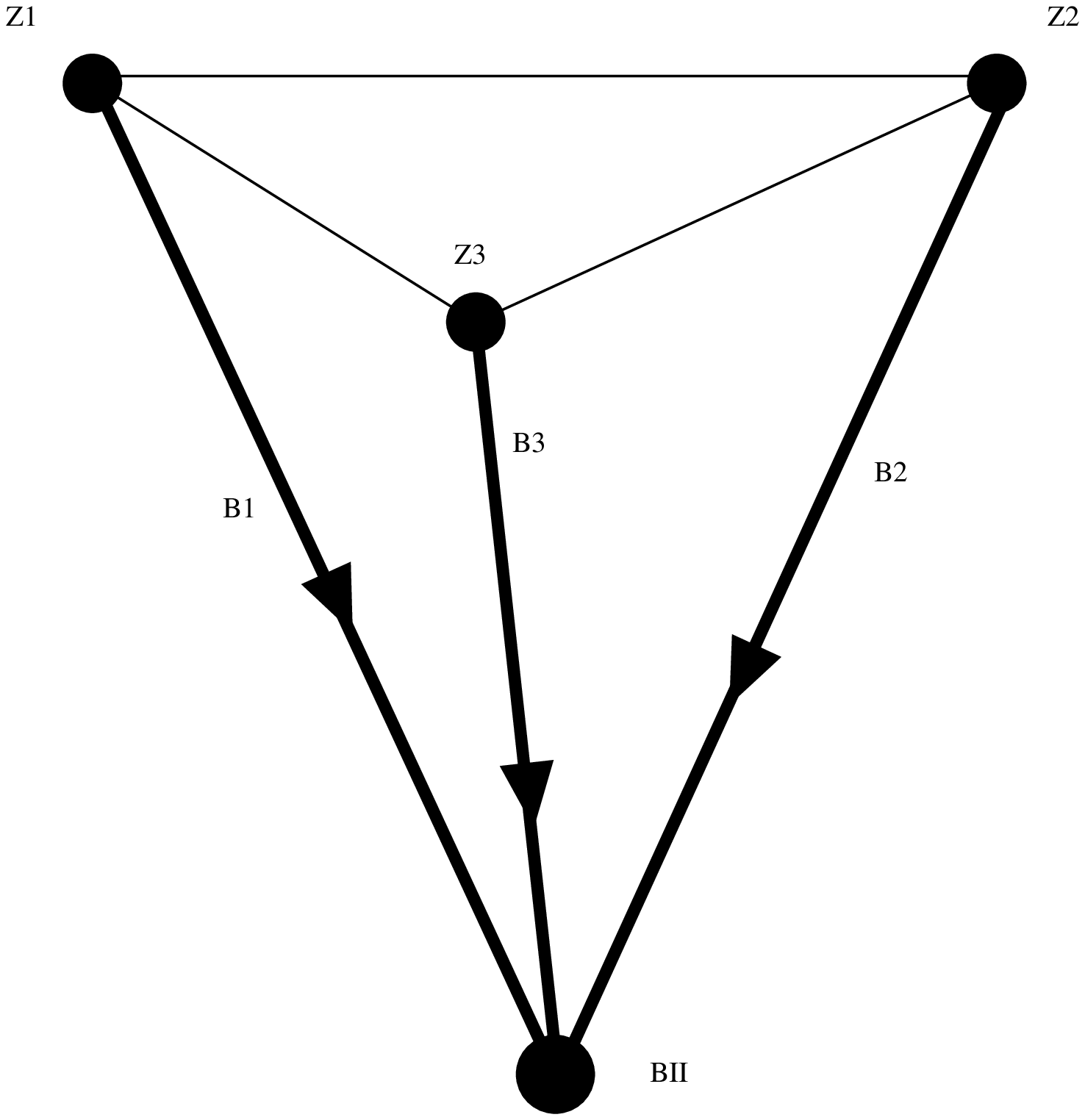}}
\caption{A simplistic picture of the state space $\overlineBIX$.
$\BVII$ consists of three `branches' ($\mathcal{B}_{\hN_1 \hN_2}$, $\mathcal{B}_{\hN_1 \hN_3}$,
$\mathcal{B}_{\hN_2 \hN_3}$)
that intersect in a common subset, $\overlineBII$.
The flow of the $\BVII$ subsets is directed away from the points
$\mathrm{Z}_\alpha$ towards $\overlineBII$.}
\label{BIXfigure}
\end{figure}

\begin{remark}
The concept of `isolated branches'---used in the proof of
Lemma~\ref{noVII} and discussed in the previous remark---is a
very useful picture to have in mind also for the general
situation. The state space $\overlineBIX$ can be depicted
roughly as the space between three branches (the three $\BVII$
subsets $\mathcal{B}_{\hN_1 \hN_2}$, $\mathcal{B}_{\hN_2
\hN_3}$, $\mathcal{B}_{\hN_3 \hN_1}$) sticking out from a
common basis (`trunk') which is the set $\overlineBII$; the end
points of the branches are the points $\mathrm{Z}_1$,
$\mathrm{Z}_2$, $\mathrm{Z}_3$; see Figure~\ref{BIXfigure}. The
main observation is that the flow on the branches is (more or
less) unidirectional, i.e., directed away from the points
$\mathrm{Z}_\alpha$ (although there are the lines
$\mathrm{TL}_\alpha$ of fixed points on $\BVII$ which make the
situation much more subtle). A generic orbit $\gamma$ cannot
have $\omega$-limit points in the interior of the state space,
see Lemma~\ref{monlemma}, however, the picture strongly
suggests that $\omega$-limit points on the branches are
excluded as well: This is simply because an orbit $\gamma$
cannot `climb up' to a point on $\BVII$ along the $\BVII$
branches themselves (which is due to the continuous dependence
of the flow on initial data). Lemma~\ref{noZandother} is
intimately connected with this idea.
\end{remark}

\begin{remark}
Lemma~\ref{noVII} implies that the $\omega$-limit set of a non-generic
orbit $\gamma$ (in the vacuum subset of $\BIX$)
is a subset of $\overlineBII \cup \{\mathrm{Z}_\alpha\}$.
It follows trivially that $\{\mathrm{Z}_\alpha\} \subsetneq \omega(\gamma)$
is impossible, which
is simply because these points are isolated
from the set $\overlineBII$ (while limit sets are
necessarily connected).
(By studying the LRS subset it is easy to show that
$\{\mathrm{Z}_\alpha\} \subseteq \omega(\gamma)$ is excluded
for LRS orbits.)
We will give an independent proof of this statement
(and its extension to the fluid case) in Lemma~\ref{noZandother}.
\end{remark}

\begin{lemma}\label{noZandother}
Let $\gamma$ be an orbit in $\BIX$.
Then $\omega(\gamma) \supsetneq \{\mathrm{Z}_1\}$ (or $\{\mathrm{Z}_2\}$, $\{\mathrm{Z}_3\}$)
is impossible.
\end{lemma}

\begin{proof}
Since the statement of the lemma is trivial for non-generic
orbits (past asymptotically self-similar orbits,
cf.~Section~\ref{nongeneric}), we restrict ourselves to generic
orbits $\gamma$. Assume that there exists a (generic) orbit
$\gamma$ such that the $\omega$-limit set of $\gamma$ contains
$\mathrm{Z}_1$, i.e., $\omega(\gamma) \supsetneq
\{\mathrm{Z}_1\}$. Accordingly, there exists a diverging
sequence of times $(\varsigma_n)_{n\in\mathbb{N}}$ such that
$\gamma(\varsigma_n) \rightarrow \mathrm{Z}_1$ as $n\rightarrow
\infty$. There exists a neighborhood $\mathscr{V}$ of
$\mathrm{Z}_1$ such that $\gamma$ intersects the complement of
$\mathscr{V}$ infinitely many times (otherwise $\gamma$ would
converge to $\mathrm{Z}_1$). Therefore we can construct a
sequence of times $(\sigma_n)_{n\in\mathbb{N}}$, $\sigma_n <
\varsigma_n$ $\forall n$, such that $\gamma(\sigma_n) \in
\partial\mathscr{V}$ and $\gamma|_{(\sigma_n,\varsigma_n]} \in
\mathscr{V}$ for all $n$. (Clearly, $|\varsigma_n-\sigma_n|$
diverges as $n\rightarrow \infty$, because
$\gamma(\varsigma_n)$ converges to the fixed point
$\mathrm{Z}_1$.) The sequence
$(\gamma(\sigma_n))_{n\in\mathbb{N}}$ possesses a converging
subsequence, i.e., there exists a point $\mathrm{P}$ such that
$\gamma(\sigma_n) \rightarrow \mathrm{P}$ as
$n\rightarrow\infty$ (where the index $n$ now runs over the
index set of the considered subsequence). By construction,
$\mathrm{P} \in \partial\mathscr{V}$; furthermore, by
definition, $\mathrm{P}$ is an $\omega$-limit point of the
orbit $\gamma$; Lemma~\ref{monlemma} implies that $\mathrm{P}
\in \BVII$ (since $\BI$ and $\BII$ are disjoint from
$\mathscr{V}$). As discussed in the subsection ``Bianchi type
$\BVII$: A new analysis,'' the $\omega$-limit of the orbit
through the point $\mathrm{P}$, which we call
$\gamma_{\mathrm{P}}$, is a fixed point on $\mathrm{TL}_1$ or
$\overlineBII$ (generically, $\omega(\gamma_{\mathrm{P}})$ is a
fixed point on the Kasner circle); in particular,
$\omega(\gamma_{\mathrm{P}})$ does not contain $\mathrm{Z}_1$.
Assume that $\omega(\gamma_{\mathrm{P}}) \in \overlineBII$;
then, by continuous dependence on initial data, for
sufficiently large $n$, $\gamma|_{(\sigma_n,\cdot)}$ shadows
the orbit $\gamma_{\mathrm{P}}$ and reaches $\overlineBII$; but
this is a contradiction to the fact that
$\gamma|_{(\sigma_n,\cdot)} \subset \mathscr{V}$. Assume that
$\omega(\gamma_{\mathrm{P}})$ is one of the fixed points of
$\mathrm{TL}_1$; if this fixed point is not contained in
$\mathscr{V}$, we immediately obtain a contradiction;
otherwise, we proceed in close analogy to the proof of
Lemma~\ref{noVII}; we exploit the (center) saddle property of
$\omega(\gamma_{\mathrm{P}})$ and we obtain that
$\gamma|_{(\sigma_n,\cdot)}$, for sufficiently large $n$,
shadows first $\gamma_{\mathrm{P}}$, then follows some orbit
emanating from $\omega(\gamma_{\mathrm{P}})$ and eventually
approaches a fixed point on the Kasner circle; this is again
the desired contradiction.
\end{proof}

\begin{lemma}[Fluid case]\label{noVIIfluid}
A generic orbit in $\BIX$ cannot have an $\omega$-limit point on the
subset $\BVII$.
\end{lemma}

\begin{proof}(\textit{Fluid case, i.e., $\hOmega > 0$.})
Consider a generic orbit $\gamma$ in $\BIX$
and assume that the
$\omega$-limit set $\omega(\gamma)$ contains a point $\mathrm{P}$ on $\BVII$.
We distinguish three possible
cases: (i) $\mathrm{P}$ is an element of the fluid subset of $\BVII$,
i.e., $\hOmega|_{\mathrm{P}} > 0$;
(ii) $\mathrm{P} \in \mathcal{QL}_\alpha$;
(iii) $\mathrm{P} \in \mathrm{TL}_\alpha$;
(iv) $\mathrm{P}$ is an element of the vacuum subset of $\BVII$,
but $\mathrm{P} \not\in \mathrm{TL}_\alpha$,
$\mathrm{P}\not\in\mathcal{QL}_\alpha$.

Consider case (i). Since $\mathrm{P} \in \omega(\gamma)$, the orbit
$\gamma_{\mathrm{P}}$ through $\mathrm{P}$ and its $\alpha$-limit
$\alpha(\gamma_{\mathrm{P}})$ must also
be contained in $\omega(\gamma)$. Our analysis of the $\BVII$ subset
shows that $\alpha(\gamma_{\mathrm{P}}) = \{\mathrm{Z}_\alpha\}$.
Consequently, $\{\mathrm{P}, \mathrm{Z}_\alpha\} \subset \omega(\gamma)$;
but this is a contradiction to Lemma~\ref{noZandother}.
Case (ii) is completely analogous.

Consider case (iii), i.e., $\omega(\gamma)$ contains a fixed point
$\mathrm{L}_\alpha$ of the line $\mathrm{TL}_\alpha$.
The scenario $\omega(\gamma) = \{\mathrm{L}_\alpha \}$
is impossible, since $\gamma$ is a generic orbit and the stable
manifold of $\mathrm{L}_\alpha$ is a subset of the LRS subset.
In analogy to the proof of Lemma~\ref{noVII} we can exploit
the (center) saddle structure of $\mathrm{L}_\alpha$, which
is reflected in~\eqref{Lalphasaddle} and the additional `fluid' equation
$\hOmega^{-1} d\hOmega/d\htau = -3(1-w) \hH$.
Hence, since $\{\mathrm{L}_\alpha\} \subsetneq \omega(\gamma)$,
it follows that $\omega(\gamma)$ contains a point $\mathrm{Q}$ on
the stable manifold of $\mathrm{L}_\alpha$; since
$\mathrm{Q} \not\in \BIX$ by Lemma~\ref{monlemma},
we have $\mathrm{Q} \in \BVII \cap \mathcal{L\!R\!S}_\alpha$,
i.e., $\mathrm{Q}$ lies on the Bianchi type~VII$_{\bm{0}}$
LRS orbit that converges to $\mathrm{L}_\alpha$.
Because $\hOmega > 0$ for this orbit (and thus for $\mathrm{Q}$),
this brings us back to case (i); a contradiction ensues.

Consider case (iv). Since the $\alpha$-limit of the orbit
through $\mathrm{P}$ is a fixed point $\mathrm{L}_\alpha$
on $\mathrm{TL}_\alpha$, case (iv) can be reduced to case (iii).
\end{proof}

\begin{lemma}\label{noII}
A generic orbit in $\BIX$ cannot have an $\omega$-limit point on the
fluid subset of $\overlineBII$.
\end{lemma}

\begin{proof}
The lemma is obviously true in the vacuum case.
In the fluid case, assume that there exists a generic orbit $\gamma$
such that $\omega(\gamma)$ contains a point $\mathrm{P} \in \BI \cup \BII$
with $\hOmega|_{\mathrm{P}} > 0$.
First, assume that $\mathrm{P}$ is an element of $\BII$,
e.g., in the $\mathcal{B}_{\hN_\alpha}$ subset.
The orbit through $\mathrm{P}$ has
the fixed point $\mathrm{CS}_\alpha$ as its $\alpha$-limit set,
see Section~\ref{subsets};
hence $\mathrm{CS}_\alpha \in \omega(\gamma)$.
Using the saddle structure of $\mathrm{CS}_\alpha$, see Section~\ref{nongeneric},
we conclude that $\omega(\gamma)$ also contains
a point $\mathrm{Q}$ of the stable manifold of $\mathrm{CS}_\alpha$.
However, since this stable manifold is a subset of $\BVII \cup \BIX$,
we have $\mathrm{Q} \in \BVII \cup \BIX$ and thus a contradiction to
Lemma~\ref{monlemma} or Lemma~\ref{noVIIfluid}.
Second, assume $\mathrm{P} \in \BI$.
Using the same line of arguments (where $\mathrm{F}$ takes
the role of $\mathrm{CS}_\alpha$) we obtain
that $\mathrm{Q} \in \BII \cup \BVII \cup \BIX$
and thus a contradiction to what has already been proved.
\end{proof}

Lemmas~\ref{noVII} and~\ref{noVIIfluid} imply
that the $\omega$-limit set of a generic orbit in $\BIX$
must be either contained in the vacuum subset of
$\overlineBII$ (where we recall that $\overlineBII = \BII \cup \BI$)
or it is a fixed point of the set
$\{\mathrm{Z}_1, \mathrm{Z}_2,\mathrm{Z}_3\}$.
It remains to prove that the latter scenario is impossible.

\begin{lemma}\label{noZ}
There does not exist any
orbit in $\BIX$
that converges to
$\mathrm{Z}_1$ ($\mathrm{Z}_2$, $\mathrm{Z}_3$)
as $\htau \rightarrow \infty$.
\end{lemma}
\begin{proof}
The non-trivial case (which is at the same time
the case that is relevant for our purposes)
is the non-LRS case.
We perform a proof by contradiction.
The main idea is to consider a non-LRS orbit
that is assumed to converge to the fixed point
$\mathrm{Z}_1$; the convergence to $\mathrm{Z}_1$ then
implies that $\hN_1$ decays
rapidly so that the orbit is forced to shadow a $\BVII$ orbit as
$\mathrm{Z}_1$ is approached.
However, there does not exist any orbit on $\BVII$ that
converges to $\mathrm{Z}_1$ as $\htau \rightarrow \infty$;
a contradiction must ensue.

This idea is formalized by using a
non-negative function $\zeta_1$ on $\overlineBIX$
which is zero at $\mathrm{Z}_1$ and whose
restriction to $\BVII$ is monotonically increasing along the flow
of $\BVII$.
Since the
orbit $\gamma$ in $\BIX$ that is assumed to converge to $\mathrm{Z}_1$
must shadow a type $\mathrm{VII}_0$ orbit,
we expect the function $\zeta_1$ to increase
along $\gamma$ as well,
or at least to decrease at a rate small
enough so that the integral still exists.
This leads directly to a contradiction, because by assumption
$\zeta_1$ must go to zero
as $\mathrm{Z}_1$ is approached.
In essence this is the main idea employed
by Ringstr\"om~\cite{rin01} in a similar context,
who, however, used $H$-normalized variables, and a
function introduced by Wainwright and Hsu~\cite{waihsu89}; to
facilitate comparison we choose basically the same function.
Consider
\begin{subequations}
\begin{align}
\zeta_1 & = \frac{(\hSigma_2 - \hSigma_3)^2 + (\hN_2 - \hN_3)^2}{\hN_2
\hN_3}\:,\\
\frac{d\zeta_1}{d\htau} & = -\frac{1}{\hN_2 \hN_3} \left[ 2
(\hSigma_2 - \hSigma_3)^2 (\hSigma_1 - 2\hH) + 2 (\hSigma_2 -
\hSigma_3) (\hN_2 - \hN_3) \hN_1 \right]\:.\label{zeta1prime}
\end{align}
\end{subequations}
The second term in the brackets can be simply estimated by using
\begin{equation}\label{ineq1}
2 (\hSigma_2 - \hSigma_3) (\hN_2 - \hN_3) \leq
(\hSigma_2 - \hSigma_3)^2 +  (\hN_2 - \hN_3)^2 = \zeta_1 \hN_2 \hN_3\:.
\end{equation}
We combine the
constraints~\eqref{Gausscon} and~\eqref{Hconstraint} to find
\begin{equation*}
\hSigma^2 - \hH^2 + \sfrac{1}{12} \left[ \hN_1^2 + (\hN_2 - \hN_3)^2
\right] + \hOmega = \sfrac{1}{6} \hN_1 (\hN_2 + \hN_3)\:.
\end{equation*}
Since $\hSigma^2 = \frac{1}{4}\hSigma_1^2 + \frac{1}{12}(\hSigma_2 -
\hSigma_3)^2$ we obtain
\begin{equation*}
\sfrac{1}{4} (\hSigma_1 - 2 \hH) (\hSigma_1 + 2\hH) + \sfrac{1}{12}
\left[ \hN_1^2 + (\hSigma_2 - \hSigma_3)^2 + (\hN_2 - \hN_3)^2
\right] + \hOmega = \sfrac{1}{6} \hN_1 (\hN_2 + \hN_3)\:.
\end{equation*}
If $\hSigma_1 -2\hH > 0$, which is the
`worst case scenario' for our considerations, then
\begin{equation}\label{ineq2}
\hSigma_1 - 2 \hH \leq \frac{1}{6} \hN_1 (\hN_2 + \hN_3)
\frac{4}{\hSigma_1 + 2 \hH} <
\frac{1}{6} (\hN_2 + \hN_3) \, \frac{\hN_1}{\hH} <
\lambda \, \frac{\hN_1}{\hH}
\end{equation}
for some sufficiently large positive
constant $\lambda$ (since $\hN_\alpha$ are bounded);
if $\hSigma_1 -2\hH \leq 0$, this inequality holds trivially.

Assume that there exists a non-LRS orbit $\gamma$
in $\BIX$ that converges to
$\mathrm{Z}_1$ as $\htau \rightarrow \infty$.
Inserting the inequalities~\eqref{ineq1} and~\eqref{ineq2}
into~\eqref{zeta1prime} yields
\begin{equation}\label{dzeta1zeta1}
\frac{d\zeta_1}{d\htau} \geq - 2 \lambda
\frac{\hN_1}{\hH} \: \zeta_1 - \hN_1 \zeta_1
\geq
-\frac{\hN_1}{\hH^2} \zeta_1\:,
\end{equation}
where the latter inequality is true for sufficiently large
$\htau$, since $\hH\rightarrow 0$ along $\gamma$.
At a reference time $\htau = \hat{\tau}$,
which we choose to be
sufficiently large (in order for~\eqref{dzeta1zeta1} to hold),
the function $\zeta_1$ takes a value $\hat{\zeta}_1 > 0$.
Integration of the
differential inequality~\eqref{dzeta1zeta1} yields
\begin{equation}\label{zetainf}
\log \zeta_1(\htau) \geq \log \hat{\zeta}_1 - \int_{\hat{\tau}}^{\htau}
\frac{\hN_1}{\hH^2}(\bar{\varsigma})\: d\bar{\varsigma}
\end{equation}
along the orbit $\gamma$.

To estimate the r.h.\ side in equation~\eqref{zetainf} we exploit
the convergence of $\gamma$ to $\mathrm{Z}_1$.
This convergence entails that $\hN_1 \rightarrow 0$ and $\hN_2
\rightarrow \sqrt{6}$, $\hN_3 \rightarrow \sqrt{6}$; the constraints
then automatically imply $\hH\rightarrow 0$, $\hSigma^2\rightarrow
0$, and $\hOmega\rightarrow 0$. Let $\delta = 2 \sqrt{6} - \hN_2
-\hN_3$. Convergence to $\mathrm{Z}_1$ implies that $\delta \rightarrow 0$.
Using the constraint~\eqref{Hconstraint}
and the fact that $\hN_1 = O(\hH^3)$
as $\hH\rightarrow 0$ (which follows from the monotonicity of $\bar{\Delta}$)
we obtain
\begin{align}
\label{Hfirst}
\hH^2 + \textfrac{2}{\sqrt{6}} \hN_1 = \textfrac{1}{\sqrt{6}}\,
\delta + o(\delta)
&\qquad\Rightarrow\qquad
\hH^2 = \textfrac{1}{\sqrt{6}}\, \delta + o(\delta) \qquad
\text{for }\, \delta\rightarrow 0\:.
\intertext{The constraint~\eqref{Gausscon} gives}
\label{secondest}
\hSigma^2 + \hOmega = \textfrac{1}{\sqrt{6}} \,\delta + o(\delta)
& \qquad\Rightarrow\qquad \frac{\hSigma^2 + \hOmega}{\hH^2} = 1 + o(1)
\qquad \text{for } \, \delta\rightarrow 0\:.
\end{align}
For the quantity
$\hq = 2 (\hSigma^2 +\hOmega) - \textfrac{3}{2}(1-w)\hOmega$
we therefore derive the estimate
$\hq/\hH^2 = 2 +o(1) - \sfrac{3}{2} (1 -w)
\hOmega/\hH^2 \geq \sfrac{1}{2} ( 1 + 3 w) + o(1)$,
while $\hq/\hH^2 = 2 + o(1)$ in the vacuum case.
We conclude that there exists
a constant $\alpha > 0$ such that
\begin{equation}\label{hqhH}
\frac{\hq}{\hH^2} \geq \frac{\alpha}{3}
\end{equation}
for sufficiently small values of $\delta$, i.e., for
sufficiently large values of $\htau$
along $\gamma$.
Using this inequality in the integrated version
of equation~\eqref{Mdecr}, i.e.,
\begin{equation}\label{Mint}
\bar{\Delta}=
\frac{\hN_1 \hN_2 \hN_3}{\hH^3}
\,\propto\,
\exp\left[ -3 \int_{\hat{\tau}}^{\htau}
\frac{\hq}{\hH} \,d\htau\right]\:,
\end{equation}
results in the estimate
\begin{equation}\label{hN1decay}
\frac{\hN_1}{\hH^3} \leq C  \exp\left[ - \alpha
\int_{\hat{\tau}}^{\htau} \hH d\htau'\right]\:,
\end{equation}
where $C$ is some positive constant and where $\htau$ is
assumed to be sufficiently large.

Therefore, $\hN_1$ goes to zero at a fast rate which
ensure finiteness of the integral in~\eqref{zetainf} as
$\htau \rightarrow \infty$:
\begin{equation}
\int_{\hat{\tau}}^\infty \frac{\hN_1}{\hH^2} d\htau =  \int_{\hat{\tau}}^\infty
\frac{\hN_1}{\hH^3} \hH d\htau  \leq C \alpha^{-1} \left[ 1 -
\exp\left( -\alpha \int_{\hat{\tau}}^\infty \hH d\htau \right) \right]
\leq C \alpha^{-1}\:.
\end{equation}
We thus conclude from~\eqref{zetainf} that $\zeta_1$ remains
bounded away from zero as $\htau \rightarrow \infty$; but this
contradicts the assumption that the orbit $\gamma$ converges to the fixed
point $\mathrm{Z}_1$, since then $\zeta_1 \rightarrow 0$ along $\gamma$.
\end{proof}

\begin{remark}
The function $\zeta_1$ can be used to give an independent proof
of the statement $\mathrm{Z}_1 \not\subset \omega(\gamma)$
$\forall \gamma$
(which is closely related to Lemma~\ref{noZandother});
we briefly sketch this proof.
Since the case $\omega(\gamma) = \{\mathrm{Z}_1\}$ is
treated in Lemma~\ref{noZ},
consider an orbit $\gamma$ such that
$\omega(\gamma) \subsetneq \{\mathrm{Z}_1\}$.
If $\gamma(\varsigma_n) \rightarrow \mathrm{Z}_1$ as
$n\rightarrow \infty$, then $\zeta_1(\varsigma_n) \rightarrow 0$
as $n\rightarrow \infty$.
Relations~\eqref{dzeta1zeta1} and~\eqref{hqhH}
hold whenever an orbit is in a sufficiently small neighborhood
of $\mathrm{Z}_1$. Therefore, in analogy to the considerations
in the proof of Lemma~\ref{noZ},
it is impossible to
achieve $\zeta_1(\varsigma_n) \rightarrow 0$,
if $\gamma$ has an $\omega$-limit point that does not lie
on the LRS subset. (In there were such a point
it would be impossible for $\zeta_1$ to decrease
sufficiently much between that point
and $\gamma(\varsigma_n)$, $n$ sufficiently large.)
A priori it is possible that
$\gamma$ has an $\omega$-limit point on the
LRS subset (note that $\zeta_1 = 0$
on the LRS subset).
However, a study of the LRS dynamics shows that orbits emanate from
$\mathrm{Z}_1$, but there do not exist orbits that converge to
$\mathrm{Z}_1$ as $\htau \rightarrow \infty$.
Using the same reasoning as above (e.g., in the remark
following Lemma~\ref{noVII} or in the proof of Lemma~\ref{noZandother})
we therefore exclude the possibility that
$\gamma$ has an $\omega$-limit point on the LRS subset as well.
Consequently, $\omega(\gamma)$ cannot have an $\omega$-limit point
except $\mathrm{Z}_1$; but this is a contradiction to the assumption.
\end{remark}

The collection of the lemmas finally yields Theorem~\ref{rinthm}
and thus completes our argument.

\textbf{Theorem~\ref{rinthm}.}
\textit{%
A generic orbit $\gamma$ in $\BIX$ has an $\omega$-limit set that is
a subset of the Mixmaster attractor, i.e.,
$\omega(\gamma) \subseteq \mathcal{A}_{\mathrm{IX}} = \BI^{\mathrm{vac.}} \cup \BII^{\mathrm{vac.}}$.}
%


\begin{proof}
Lemma~\ref{monlemma} implies that the $\omega$-limit set of
a generic orbit in $\BIX$ is a subset of $\overlineBVII$,
which is given by~\eqref{BVIIsplit}.
Lemmas~\ref{noVII} and~\ref{noVIIfluid} exclude $\BVII$,
Lemma~\ref{noII} excludes the fluid subsets of $\BI$ and $\BII$.
Lemma~\ref{noZ} (in combination with Lemma~\ref{noZandother}) excludes
$\{\mathrm{Z}_1,\mathrm{Z}_2,\mathrm{Z}_3\}$.
This leaves the vacuum subset of $\BI \cup \BII$ as the only possible
superset of the $\omega$-limit set of a generic orbit.
\end{proof}

\begin{remark}
To complete the statement of Theorem~\ref{rinthm}
it is important to note that
both $\omega(\gamma) \subseteq \BI^{\mathrm{vac.}}$
and $\omega(\gamma) \subseteq  \BII^{\mathrm{vac.}}$
is impossible. The latter is obvious, since
$\BII^{\mathrm{vac.}}$ consists of a collection of
heteroclinic orbits (transitions) with end points
on $\BI^{\mathrm{vac.}}$ ($= \mathrm{K}^\ocircle$),
see Section~\ref{subsets}.
The proof that
$\omega(\gamma) \subseteq \BI^{\mathrm{vac.}}$
($= \mathrm{K}^\ocircle$) is impossible
is contained in the local analysis
of Section~\ref{nongeneric}.
(For an alternative proof see~\cite{rin00}.)
\end{remark}

\begin{remark}
It is important to emphasize that the Theorem states
that $\omega(\gamma) \subseteq \mathcal{A}_{\mathrm{IX}}$.
Whether $\omega(\gamma)$ actually  coincides with $\mathcal{A}_{\mathrm{IX}}$
(at least generically)
or whether it is a proper subset of $\mathcal{A}_{\mathrm{IX}}$ is open.
This question and related issues are discussed in detail
in~\cite{heiuggpreprint}.
\end{remark}

\section{Consequences}
\label{consequences}

The Bianchi type~IX attractor theorem
in conjunction with our understanding of the flow on the attractor
subset
implies a number of consequences 
that we formulate as corollaries.
Some of these correspond to results presented in~\cite{rin01};
our approach, however, is rather different.

\begin{figure}[ht]
\psfrag{a}[cc][cc]{$\Sigma_1$} \psfrag{b}[cc][cc]{$\Sigma_2$}
\psfrag{c}[cc][cc]{$\Sigma_3$} \psfrag{k}[cc][cc]{$\mathrm{T}_1$}
\psfrag{l}[cc][cc]{$\mathrm{T}_2$}
\psfrag{m}[cc][cc]{$\mathrm{T}_3$}
\psfrag{n}[cc][cc]{$\mathrm{Q}_1$}
\psfrag{o}[cc][cc]{$\mathrm{Q}_2$}
\psfrag{p}[cc][cc]{$\mathrm{Q}_3$}
\centering
\includegraphics[height=0.35\textwidth]{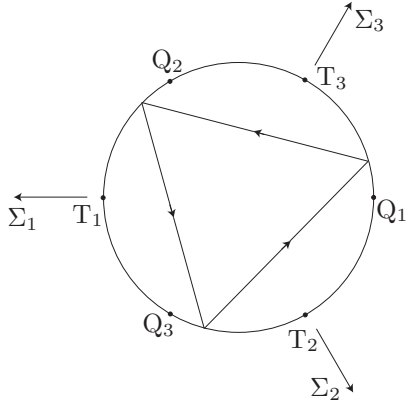}
  \caption{The `simplest' structure on the Mixmaster attractor:
    a heteroclinic cycle with three fixed points. In general, concatenating of
    the vacuum type~II orbits of Figure~\ref{alltypeII} yields infinite heteroclinic
    chains.}
\label{period1a}
\end{figure}

On the Mixmaster attractor $\mathcal{A}_{\mathrm{IX}}$,
the dynamical system~\eqref{domsys} generates an intricate
network of structures that are invariant under the flow:
heteroclinic cycles and finite
and infinite heteroclinic chains.
(These heteroclinic structures arise by concatenating
the vacuum type~II orbits on $\mathcal{B}_{\hN_1}$, $\mathcal{B}_{\hN_2}$,
and $\mathcal{B}_{\hN_3}$, see Figure~\ref{alltypeII}.)
The `simplest' structure (i.e., the structure that
contains the smallest number of fixed points)
is a heteroclinic cycle with
three fixed points on $\mathrm{K}^\ocircle$, see Figure~\ref{period1a}.
We refer to~\cite{heiuggpreprint} and references therein
for a comprehensive discussion.

\begin{lemma}\label{onethenall}
If\/ $\mathrm{P}\in \mathcal{A}_{\mathrm{IX}}$ is an $\omega$-limit
point of a type~IX orbit, then the entire heteroclinic cycle/chain 
through\/ $\mathrm{P}$
must be contained in the $\omega$-limit set.
\end{lemma}

\begin{proof}
The lemma follows from basic facts of the theory of dynamical systems~\cite{perko}.
\end{proof}

Lemma~\ref{onethenall} entails that the heteroclinic cycles and chains are
potential limit set candidates for generic type~IX orbits.

\begin{corollary}
The asymptotic behavior of a generic Bianchi type~IX solution
is oscillatory with oscillations between at least three
fixed points on $\mathrm{K}^\ocircle$.
\end{corollary}

\begin{proof}
The simplest structure on $\mathcal{A}_{\mathrm{IX}}$
that is a potential $\omega$-limit set for a generic type~IX
orbit is a heteroclinic cycle with three fixed points on $\mathrm{K}^\ocircle$,
see Figure~\ref{period1a}.
A type~IX orbit converging to such a
heteroclinic cycle (if such an orbit exists)
exhibits oscillations between three Kasner points.
\end{proof}

\begin{remark}
The oscillatory behavior implies that generic asymptotic
type~IX dynamics constitute an example of asymptotic
self-similarity breaking~\cite{limetal06}.
\end{remark}

\begin{corollary}\label{taubthenmany}
If one of the Taub points $\{\mathrm{T}_1, \mathrm{T}_2,
\mathrm{T}_3\}$ is an $\omega$-limit point of a type~IX orbit,
then the $\omega$-limit set contains Kasner fixed points
arbitrarily close to the Taub points.
\end{corollary}

\begin{proof}
Assume the contrary, i.e.,
suppose that there exists a type~IX orbit $\gamma$ such that
$\mathrm{T}_\alpha$ (for some $\alpha$)
is an element of $\omega(\gamma)$, while
at the same time there exists a neighborhood $\mathcal{U}$ of $\mathrm{T}_\alpha$
such that $\omega(\gamma)\cap \mathcal{U} \cap \mathrm{K}^\ocircle = \{\mathrm{T}_\alpha\}$.
However,
$\omega(\gamma) \cap \mathcal{U} \supsetneq \{\mathrm{T}_\alpha\}$, since
$\omega(\gamma)$ is connected and strictly larger than $\mathrm{T}_\alpha$.
(There do no exist type~IX orbits that converge to a Taub point, see 
Section~\ref{nongeneric}.)
Taking into account the structure of orbits on $\mathcal{A}_{\mathrm{IX}}$ we
conclude that
$\omega(\gamma)\cap \mathcal{U} = \bar{\varsigma} \cap\mathcal{U}$,
where $\bar{\varsigma}$ denotes the closure of the type~II 
orbit
$\mathrm{Q}_\alpha\rightarrow\mathrm{T}_\alpha$.
However, since $\omega$-limit sets cannot contain `isolated branches'
of this type, we obtain a contradiction.
\end{proof}

\begin{remark}
Corollary~\ref{taubthenmany} implies that the $\omega$-limit set contains
an infinite set of Kasner fixed points in a neighborhood of the Taub
point(s), but this set is not necessarily a continuum of
fixed points. 
\end{remark}



\begin{lemma}\label{doesnotgoawayLemma}
Let $\mathcal{W}$ be a neighborhood of the Mixmaster attractor $\mathcal{A}_{\mathrm{IX}}$.
Then there exists a smaller neighborhood $\mathcal{V}$ of $\mathcal{A}_{\mathrm{IX}}$,
$\mathcal{V} \subset \mathcal{W}$, such that each solution
with initial data in $\mathcal{V}$ (at $\htau = 0$)
remains in $\mathcal{W}$ for all $\htau > 0$.
\end{lemma}

\begin{proof}
Henceforth, we denote by $X(\mathring{x};\htau)$
the type~IX solution generated by initial data $\mathring{x} \in \BIX$, i.e.,
$X(\mathring{x},\htau)$ is a type~IX solution and $X(\mathring{x},0)= \mathring{x}$.
We employ the two functions $\hOmega$ and $\hN_1 \hN_2 + \hN_1 \hN_3 + \hN_2 \hN_3$, cf.~\eqref{prevform},
as a measure of the distance from the Mixmaster attractor.
The choice of a small neighborhood of the Mixmaster attractor
then corresponds to both $\hOmega$ and $\hN_1 \hN_2 + \hN_1 \hN_3 + \hN_2 \hN_3$
being bounded by a small constant $\epsilon$.
Choose a small neighborhood $\mathcal{W}$ of $\mathcal{A}_{\mathrm{IX}}$
(corresponding to a choice $\epsilon_{\mathcal{W}}$) and
consider a neighborhood $\mathcal{V}$ of $\mathcal{A}_{\mathrm{IX}}$
with $\mathcal{V} \subset \mathcal{W}$, $\epsilon_{\mathcal{V}} \ll \epsilon_{\mathcal{W}}$.
A solution $X(\mathring{x},\htau)$
with initial data 
$\mathring{x}\in \mathcal{V}$ remains in $\mathcal{W}$ at least
for $0 \leq \htau < \hT$. The constant $\hT$ depends on
$\epsilon_{\mathcal{V}}$ and, a priori, on the initial data; we
intend to show that $\hT = \infty$ irrespective of the choice
of initial data if $\epsilon_{\mathcal{V}}$ is sufficiently
small.

First, consider Eq.~\eqref{hOmegaeq} for $\hOmega$. At the
Kasner circle we obtain $\hOmega^{-1}
d\hOmega/d\htau\,|_{\mathrm{K}^{\ocircle}} = -3(1-w)$; hence
there exists a neighborhood $\mathcal{U}_{\mathrm{I}}$ of
$\mathrm{K}^\ocircle$ such that $\hOmega$ is (exponentially)
decreasing as long as the solution stays
$\mathcal{U}_{\mathrm{I}}$. In general, we obtain $\hOmega^{-1}
d\hOmega/d\htau < (1 + 3 w)$ in $\mathcal{W}$, i.e., a limit on
the possible increase. Decompose $\mathcal{W}$ into
$\mathcal{W}_{\mathrm{I}} = \mathcal{W} \cap
\mathcal{U}_{\mathrm{I}}$ and $\mathcal{W}_{\mathrm{II}} =
\mathcal{W}\backslash \mathcal{U}_{\mathrm{I}}$. Using the flow
on $\mathcal{A}_{\mathrm{IX}}$ (and elementary results from the
theory of dynamical systems) we see that the solution
$X(\mathring{x},\htau)$ oscillates between
$\mathcal{W}_{\mathrm{I}}$ and $\mathcal{W}_{\mathrm{II}}$,
where the sojourn times in $\mathcal{W}_{\mathrm{I}}$ are large
compared to the sojourn times in $\mathcal{W}_{\mathrm{II}}$.
(The ratio of the respective sojourn times diverges with
$\epsilon_{\mathcal{V}} \rightarrow 0$.) Using the estimates
for $d\hOmega/d\tau_-$ it follows that $\hOmega$ exhibits a
(rapid) overall decrease for $\htau \in [0,\hT)$ (i.e., as long
as $X(\mathring{x},\htau) \in \mathcal{W}$). Consequently, at
$\htau = \hT$, the quantity $\hOmega$ is smaller than
initially, $\hOmega|_{\hT} < \epsilon_{\mathcal{V}}$.

Second, consider $\hN := \hN_1 \hN_2 + \hN_1 \hN_3 + \hN_2
\hN_3$. For $(\alpha\beta\gamma) \in \{(123),(231),(312)\}$ we
obtain $(\hN_\alpha \hN_\beta)^{-1} \,d (\hN_\alpha
\hN_\beta)/d\htau = -2 (\hq \hH - \hSigma_\gamma + \hF)\,$
from~\eqref{Nalpha}. Let $\mathcal{W}_{\mathrm{T}}$ be a
(small) neighborhood of the Taub points.
Then $\hN_\alpha \hN_\beta$ and thus $\hN$ 
is (exponentially) decreasing in $\mathcal{W}_{\mathrm{I}} \backslash \mathcal{W}_{\mathrm{T}}$.
In $\mathcal{W}_{\mathrm{II}}$ we obtain 
$\hN^{-1} d\hN/d\htau < \mathrm{const}$,
i.e., a limit on the possible increase.
To obtain information about $\hN$ 
in $\mathcal{W}_{\mathrm{T}}$
we use the center manifold reduction theorem:
The set $\mathcal{W}_{\mathrm{T}}$ contains a piece of the Taub line $\mathrm{TL}_\gamma$
(up to some value of $\hN_\alpha = \hN_\beta$).
The analysis of $\overlineBVII^{\mathrm{vac.}}$ in Section~\ref{Bianchisevenzero}
implies that each fixed point $\mathrm{L}_\gamma$ on $\mathrm{TL}_\gamma$ is the (non-hyperbolic)
source for a one-parameter family of orbits in $\overlineBVII^{\mathrm{vac.}}$.
The family of orbits emerging from $\mathrm{L}_\gamma$
forms a two-dimensional invariant surface $\mathcal{S}[\mathrm{L}_\gamma]$
in $\overlineBVII^{\mathrm{vac.}}\cap \mathcal{W}_{\mathrm{T}}$.
The proof of Lemma~\ref{sevenlemma} entails that there
exists a universal constant $c_{\mathrm{T}}$ such that
\begin{equation}\label{variationN}
\sup\nolimits_{\mathcal{S}[\mathrm{L}_\gamma]} (\hN_\alpha \hN_\beta) \leq
c_{\mathrm{T}}\,
\inf\nolimits_{\mathcal{S}[\mathrm{L}_\gamma]} (\hN_\alpha \hN_\beta)
\end{equation}
for each surface $\mathcal{S}[\mathrm{L}_\gamma]$
(in fact, $c_{\mathrm{T}}$ is only marginally larger than $1$).
We thus obtain, 
in $\overlineBVII^{\mathrm{vac.}}\cap \mathcal{W}_{\mathrm{T}}$,
a foliation
of $\overlineBVII^{\mathrm{vac.}}$ into two-dimensional invariant surfaces
across which the relative variation of $(\hN_\alpha \hN_\beta)$ is small.
Since each point $\mathrm{L}_\gamma$ is a center saddle with a
two-dimensional stable subspace/manifold, the foliation of
$\overlineBVII^{\mathrm{vac.}}\cap \mathcal{W}_{\mathrm{T}}$
carries over, by the center manifold theorem, to a foliation of
$\mathcal{W}_{\mathrm{T}}$ into four-dimensional invariant
hypersurfaces $\mathcal{H}[\mathrm{L}_\gamma]$. (Each
hypersurface is associated with the direct sum of the stable
subspace and the tangent space of
$\mathcal{S}[\mathrm{L}_\gamma]$ at $\mathrm{L}_\gamma$.)
Furthermore, for sufficiently small $\mathcal{W}_{\mathrm{T}}$, Eq.~\eqref{variationN}
carries over from $\mathcal{S}[\mathrm{L}_\gamma]$ to $\mathcal{H}[\mathrm{L}_\gamma]$.
Therefore, since every solution in $\mathcal{W}_{\mathrm{T}}$
is contained in one of the hypersurfaces $\mathcal{H}[\mathrm{L}_\gamma]$,
we obtain $(\hN_\alpha \hN_\beta)(\htau) \leq c_{\mathrm{T}} (\hN_\alpha \hN_\beta)(\hsigma)$
for all $\htau$ such that the solution is contained in $\mathcal{W}_{\mathrm{T}}$,
where $\hsigma$ is the time the solution enters $\mathcal{W}_{\mathrm{T}}$.
(Note in particular that this result is independent of the actual time the
solution spends in the neighborhood $\mathcal{W}_{\mathrm{T}}$.)
Summing up: For the solution $X(\mathring{x},\htau)$ we find
a decrease of $\hN$ 
in $\mathcal{W}_{\mathrm{I}} \backslash \mathcal{W}_{\mathrm{T}}$
(comparatively long sojourn times), an increase in $\mathcal{W}_{\mathrm{II}}$
(comparatively short sojourn times) and a bound on the increase
in $\mathcal{W}_{\mathrm{T}}$ (irrespective of the sojourn times).
From these facts we infer an overall decrease of $\hN$
for $\htau \in [0,\hT)$ (i.e., as long as $X(\mathring{x},\htau) \in \mathcal{W}$).
Consequently, at $\htau = \hT$, the quantity $\hN$
is smaller than initially, $\hN|_{\hT} < \epsilon_{\mathcal{V}}$.

Since both $\hOmega|_{\hT} < \epsilon_{\mathcal{V}} < \epsilon_{\mathcal{W}}$ and
$\hN|_{\hT} < \epsilon_{\mathcal{V}} < \epsilon_{\mathcal{W}}$,
the solution $X(\mathring{x},\htau)$ does not actually leave $\mathcal{W}$ at
$\hT$ but is still in $\mathcal{V}$ ($\subset \mathcal{W}$) for some time beyond $\hT$.
It is immediate that this leads to $\hT = \infty$, and hence the lemma is established.
\end{proof}

\begin{remark}
Note that not all solutions in $\mathcal{V}$ or $\mathcal{W}$ are generic.
In these neighborhoods of $\mathcal{A}_{\mathrm{IX}}$,
there also exist LRS solutions, which converge
to $\mathrm{TL}_\alpha$, $\alpha =1,2,3$, instead of to
$\mathcal{A}_{\mathrm{IX}}$.
\end{remark}

\begin{corollary}\label{uniformcor}
Convergence to the Mixmaster attractor is uniform on compact sets of
generic initial data:
Let $\mathscr{X}$ be a compact set in $\BIX$ that does not intersect any of the
manifolds $\mathscr{F}$, $\mathcal{C\!S}_\alpha$, $\mathcal{L\!R\!S}_\alpha$,
so that each initial data $\mathring{x} \in \mathscr{X}$ generates a generic
type~IX solution. 
Let $X(\mathring{x};\htau)$ denote the type~IX solution with $X(\mathring{x},0) = \mathring{x}$.
Then
\begin{equation}
\| X(\mathring{x}; \htau) - \mathcal{A}_{\mathrm{IX}} \| \rightarrow 0 \qquad (\htau\rightarrow \infty)
\end{equation}
uniformly in $\mathring{x} \in \mathscr{X}$.
\end{corollary}


\begin{proof}
Let $\mathcal{W}$ be a neighborhood of the Mixmaster attractor $\mathcal{A}_{\mathrm{IX}}$.
We need to show that there exists $\hT$ such that
$X(\mathring{x};\htau) \in \mathcal{W}$ for all $\htau \geq \hT$,
for all $\mathring{x} \in\mathscr{X}$.
Let $\mathcal{V}$ be as in Lemma~\ref{doesnotgoawayLemma}.
Theorem~\ref{rinthm} implies that
for each $\mathring{x} \in \mathscr{X}$ there exists $\hsigma(\mathring{x})$ such
that $X(\mathring{x};\htau) \in \mathcal{V}$.
Since $\mathscr{X}$ is a compact set of generic initial data,
$\hT:= \sup_{\mathring{x}\in \mathscr{X}}\hsigma(\mathring{x})$ exists.
(If this supremum did not exist, we could consider a sequence
of initial data along which $\hsigma$ diverges to
construct generic initial data violating Theorem~\ref{rinthm}.)
Applying Lemma~\ref{doesnotgoawayLemma} we conclude
that $X(\mathring{x};\htau) \in \mathcal{W}$ for all $\htau \geq \hT$,
for all $\mathring{x} \in\mathscr{X}$.
\end{proof}

\begin{corollary}
For generic solutions of Bianchi type~IX the Weyl curvature scalar
$C_{abcd}C^{abcd}$ (and therefore also the Kretschmann scalar)
becomes unbounded towards the past.
\end{corollary}

\begin{proof}
At a fixed point on $\mathrm{K}^\ocircle$ (which represents a Kasner
solution) the Hubble-normalized Weyl curvature scalar
$C_{abcd}C^{abcd}/(48 H^4)$ is given by $27(2
-\Sigma_1\Sigma_2\Sigma_3)$. 
Therefore,
$C_{abcd}C^{abcd}/(48 H^4) \in [0,4]$ on the Kasner circle $\mathrm{K}^\ocircle$,
where $C_{abcd}C^{abcd}/(48 H^4) = 0$ holds only at the Taub points
$\{\mathrm{T}_1,\mathrm{T}_2,\mathrm{T}_3\}$.
Since the $\omega$-limit set of a generic
type~IX solution must necessarily contain a fixed point on
$\mathrm{K}^\ocircle$ different from the Taub points
by Corollary~\ref{taubthenmany}, we conclude
that $C_{abcd}C^{abcd}$ becomes unbounded towards the past; we simply
use that $H \rightarrow \infty$ as $\htau\rightarrow \infty$ ($t\rightarrow 0$).
\end{proof}

\begin{corollary}
Taking into account both the expanding and contracting phases of
Bianchi type~IX solutions, generic Bianchi type~IX
initial data generate an inextendible maximally
globally hyperbolic development associated with past
and future singularities where the curvature becomes unbounded.
\end{corollary}

\begin{remark}
This is a direct consequence of the previous corollary.
It follows straightforwardly that the analogous statement
holds for the asymptotically self-similar solutions as well,
the only exceptions being the type IX vacuum LRS solutions.
\end{remark}


\section{Discussion}
\label{concl}

In this paper we give new and comparatively short proofs of the
main rigorous results on Bianchi type~IX asymptotic dynamics:
Ringstr\"om's Bianchi type~IX attractor theorem~\cite{rin01},
Theorem~\ref{rinthm}, and its consequences. To find more
succinct arguments is not our primary motivation to
re-investigate the problem. By emphasizing the importance of
the Lie contraction hierarchy our proof demonstrates that
Bianchi type~IX is special in comparison with the other
oscillatory Bianchi types: type~VIII and
type~$\mathrm{VI}_{-1/9}$. Let us elaborate.

Among the class~A Bianchi models Bianchi type~IX is
characterized by the condition that the three structure
constants possess the same
sign, 
see~Table~\ref{classAmodels}; in terms of the dynamical systems
variables of Section~\ref{basic} positivity of the structure
constants is expressed by the conditions $\hN_1 > 0$, $\hN_2 >
0$, $\hN_3 > 0$. The pivotal feature of type~IX dynamics is the
following fact: The set of asymptotic states that are
accessible to Bianchi type~IX models in the past asymptotic
limit is represented by the `Bianchi type~IX Lie contraction
hierarchy' of Figure~\ref{contraction}. Since this hierarchy is
obtained by successively setting the type~IX structure
constants to zero (Lie contractions), the first level is taken
by the three equivalent representations of the Bianchi
type~$\mathrm{VII}_0$ state space, the second level consists of
the three representation of Bianchi type~II, and the third
level coincides with Bianchi type~I, see
Figure~\ref{contraction}. Accordingly, admissible past
asymptotic states of type~IX models are of Bianchi type~I,
where $\hN_1 = 0$, $\hN_2 = 0$, $\hN_3 =0$, of Bianchi type~II,
where $\hN_\alpha >0$, $\hN_\beta = \hN_\gamma = 0$, or of
Bianchi type~$\mathrm{VII}_0$, where $\hN_\alpha >0$,
$\hN_\beta >0$, $\hN_\gamma = 0$; cf.~Figure~\ref{contraction}.
(As usual, $(\alpha\beta\gamma)$ runs over the set
$\left\{(123),(231),(312)\right\}$.) Models of Bianchi
type~$\mathrm{VI}_0$ are not among the admissible past
asymptotic states, since type~$\mathrm{VI}_0$ does not appear
in the Bianchi type~IX Lie contraction hierarchy; this is
because one of the structure constants is necessarily negative
in type~$\mathrm{VI}_0$. (Using dynamical systems terminology
to summarize: The $\omega$-limit set of every type~IX orbit
lies on the boundary of the type~IX state space, which is the
union of the type~I, type~II, and type~$\mathrm{VII}_0$ state
spaces; in contrast, the type~$\mathrm{VI}_0$ state space is
not part of the boundary.)

For simplicity and clarity, let us restrict our discussion
to the vacuum case.
Vacuum Bianchi type~I solutions (Kasner solutions) are
represented by fixed points on the Kasner circle,
$\mathrm{K}^{\ocircle}$, while vacuum Bianchi type~II solutions
are represented by heteroclinic orbits that connect one fixed
point on $\mathrm{K}^{\ocircle}$ with another; see
Figure~\ref{alltypeII}. The Mixmaster attractor (which is
simply the union of the Kasner circle and the three equivalent
representations of the type~II vacuum subset) is covered by an
intricate network of these heteroclinic orbits, which can be
concatenated to form heteroclinic cycles and chains. While
Bianchi type~I and type~II thus fit together seamlessly to form
the fabric of type~IX asymptotics, Bianchi
type~$\mathrm{VII}_0$ is the odd one out. Solutions of
type~$\mathrm{VII}_0$ do not connect one fixed point on the
Kasner circle with another, but connect a fixed point on
$\mathrm{K}^{\ocircle}$ with a fixed point on the Taub line,
for which $\hN_\alpha = \hN_\beta > 0$. As a consequence,
Bianchi type~$\mathrm{VII}_0$ orbits are incompatible with the
network of heteroclinic cycles/chains; in particular, Bianchi
type~$\mathrm{VII}_0$ orbits cannot be concatenated with
Bianchi type~II orbits to form heteroclinic chains. This
incompatibility is crucial: It can be regarded as the
underlying reason that excludes Bianchi type~$\mathrm{VII}_0$
from the set of possible past asymptotic states. Indeed, the
new proof of Theorem~\ref{rinthm} given in this paper reflects
this idea accurately.

\begin{figure}[ht]
\psfrag{a}[cc][cc]{$\mathcal{B}_{\hN_1^- \hN_2^+\hN_3^+}$}
\psfrag{b}[cc][cc]{$\mathcal{B}_{\hN_1^- \hN_2^+}$}
\psfrag{c}[cc][cc]{$\mathcal{B}_{\hN_2^+ \hN_3^+}$}
\psfrag{d}[cc][cc]{$\mathcal{B}_{\hN_1^- \hN_3^+}$}
\psfrag{e}[cc][cc]{$\mathcal{B}_{\hN_1^-}$}
\psfrag{f}[cc][cc]{$\mathcal{B}_{\hN_2^+}$}
\psfrag{g}[cc][cc]{$\mathcal{B}_{\hN_3^+}$}
\psfrag{h}[cc][cc]{$\mathcal{B}_\emptyset$}
\psfrag{i}[rc][cc]{$\BVIII$}
\psfrag{j}[rc][cc]{$\BVI/\BVII$}
\psfrag{k}[rc][cc]{$\BII$}
\psfrag{l}[rc][cc]{$\BI$} \centering{
\includegraphics[height=0.45\textwidth]{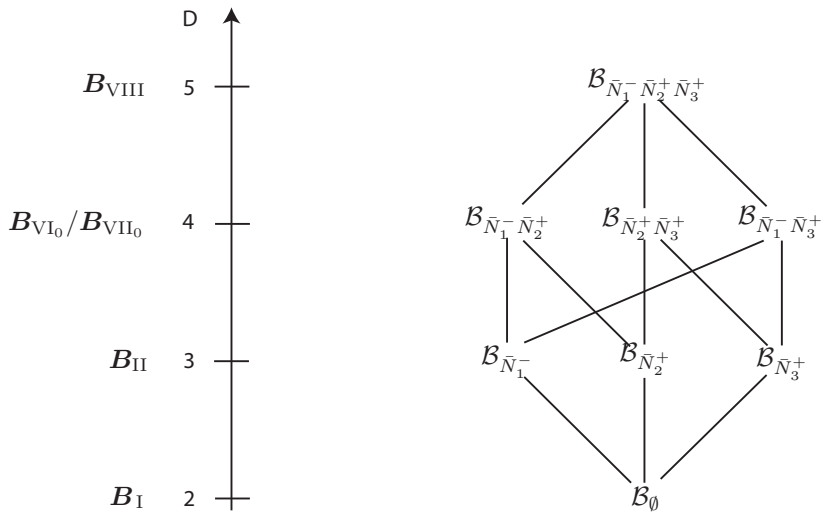}}
\caption{This subset contraction diagram for Bianchi type~VIII
is to be compared with the one of type~IX, see Figure~\ref{contraction}.
The subscript of $\mathcal{B}_\star$ denotes the non-zero
variables; variables with the superscript ${}^+$ are
are positive, variables with the superscript ${}^-$ are negative;
e.g. $\mathcal{B}_{\hN_1^-\hN_2^+}$ denotes the type
$\mathrm{VI}_0$ subset with $\hN_1 < 0$, $\hN_2 >0$ and $\hN_3 = 0$.
Note that $\mathcal{B}_{\hN_1^-}$ is a completely equivalent representation of
$\mathcal{B}_{\hN_1^+}$.}
\label{contraction2}
\end{figure}

The exclusion of Bianchi type~$\mathrm{VII}_0$ from the set of
admissible past asymptotic states is the cornerstone of the
analysis of type~IX asymptotics. Once established, the
exclusion of type~$\mathrm{VII}_0$ automatically leaves the
Mixmaster attractor (or a subset thereof) as the past attractor
and thus yields Theorem~\ref{rinthm}. So what about Bianchi
type~VIII then? There is but one difference between type~VIII
and type~IX that is relevant in the context of past asymptotic
dynamics: The Lie contraction hierarchies of Bianchi type~VIII
and Bianchi type~IX differ on the first level, compare
Figure~\ref{contraction} with Figure~\ref{contraction2}.
The boundary of the type~VIII state space encompasses a Bianchi
type~$\mathrm{VII}_0$ state space and two equivalent
representations of the Bianchi type~$\mathrm{VI}_0$ state space
instead of three type~$\mathrm{VII}_0$ representations; hence
the Bianchi type~VIII Lie contraction hierarchy contains a
representation of each class~A model, whereas Bianchi type~IX
does not. (This is merely one aspect of Bianchi type~VIII being
more general than type~IX. Another aspect is the violation of
the permutation symmetry in type~VIII.)

Bianchi type~$\mathrm{VI}_0$ and Bianchi type~$\mathrm{VII}_0$
possess the same scale-automorphism group which generates
corresponding monotone functions~\cite{waiell97}. However, the
state spaces on which these monotone functions act are rather
different: While the type~$\mathrm{VII}_0$ state space is
unbounded and contains a line of fixed points, the
type~$\mathrm{VI}_0$ state space is bounded and does not
contain this line of fixed points. The reason for these
differences between type~$\mathrm{VI}_0$ and $\mathrm{VII}_0$
is that type~$\mathrm{VI}_0$ does not admit an LRS subset,
while $\mathrm{VII}_0$ does. In type~$\mathrm{VII}_0$ the
vacuum LRS subset contains a one-parameter set of
representations of Minkowski spacetime which are different, but
equivalent, from the Taub representation in Bianchi type~I. In
the dynamical systems picture this corresponds to a line of
fixed points, $\mathrm{TL}_\alpha$, which is necessarily absent
in type~$\mathrm{VI}_0$. These features indicate that type
$\mathrm{VI}_0$ dynamics is simpler than that of type
$\mathrm{VII}_0$, which is indeed the case, see,
e.g.,~\cite{heietal07,waiell97,waietal99}. Paradoxically, the
very simplicity of the type~$\mathrm{VI}_0$ flow is the reason
why the analysis of Bianchi type~VIII asymptotics is more
intricate and complex than that of type~IX. Let us elaborate.
In contrast to Bianchi type~$\mathrm{VII}_0$ solutions, orbits
of Bianchi type~$\mathrm{VI}_0$ are compatible with the network
of heteroclinic chains on the Mixmaster attractor. This is a
crucial fact, because it implies that Bianchi
type~$\mathrm{VI}_0$ orbits can be concatenated
with heteroclinic chains. 
Therefore, Bianchi type~$\mathrm{VI}_0$ states cannot be ruled
out a priori as possible asymptotic states of Bianchi type VIII
solutions. It is clear that this causes the (analog of the)
proofs of Theorem~\ref{rinthm} to fail in type~VIII. However,
the failure of the methods of proof does not imply the failure
of the statement: It is expected that (a generic version of)
Theorem~\ref{rinthm} holds and that the Mixmaster attractor is
the past attractor for type~VIII models. However, it is
conceivable that arguments of a completely different kind, like
stochastic arguments, are necessary to prove that Bianchi
type~$\mathrm{VI}_0$ is excluded (generically) from being
involved in the asymptotic dynamics of solutions. A statement
like this would then be the core of a (generic) version of
Theorem~\ref{rinthm} for Bianchi type~VIII.

A closely related issue concerns the validity of
Lemma~\ref{doesnotgoawayLemma} and the related
Corollary~\ref{uniformcor}. Lemma~\ref{doesnotgoawayLemma}
fails in Bianchi type~VIII. There exists (generic) initial data
that is arbitrarily close to the Mixmaster attractor but
transported beyond a given neighborhood of the attractor by the
flow of the dynamical system.
The reason for this difference
between type~VIII and~IX is the by now familiar one: The
contraction hierarchy of type~VIII contains type
$\mathrm{VI}_0$ while that of type~IX does not.
However, lack of uniform convergence
does not imply that there is no convergence at all in type~VIII:
The Mixmaster attractor might still be the past attractor.

Our arguments support the thesis that Bianchi type~IX is
special. The existence of discrete symmetries associated with
axes permutations, which follow from the positivity of the
structure constants, leads to a simplification of the problem
and makes the treatment of type~IX dynamics relatively
straightforward. Type~VIII on the other hand is less symmetric
and is thus harder to grasp. The same is true for the other
remaining oscillatory Bianchi models---the Bianchi
type~$\mathrm{VI}_{-1/9}$ models. These types might therefore
be more relevant for our understanding of generic spacelike
singularities.

\subsection*{Acknowledgments}
We thank Alan Rendall, Hans Ringstr\"om, and especially Lars Andersson
for useful discussions.
We gratefully acknowledge the hospitality of the Mittag-Leffler Institute,
where part of this work was completed.
CU is supported by the Swedish Research Council.

\bibliographystyle{plain}

\begin{thebibliography}{90}










\bibitem{waiell97}
J. Wainwright and G.F.R. Ellis.
\newblock {\em Dynamical systems in cosmology}.
\newblock (Cambridge University Press, Cambridge, 1997).












\bibitem{ren97}
A.D. Rendall.
\newblock Global dynamics of the mixmaster model.
\newblock {\it Class. Quantum Grav.} {\bf 14} 2341 (1997). 

\bibitem{rin00}
H. Ringstr\"om.
\newblock Curvature blow up in Bianchi VIII and IX vacuum
spacetimes.
\newblock {\it Class.\ Quantum Grav.} {\bf 17} 713 (2000). 

\bibitem{rin01}
H. Ringstr\"om.
\newblock The Bianchi IX attractor.
\newblock {\it Annales Henri Poincar\'e} {\bf 2} 405 (2001).

\bibitem{perko} L. Perko.
\newblock {\em Differential Equations and Dynamical Systems.}
\newblock (Springer, New York, 2001).

\bibitem{cra91} J.~D.~Crawford. Introduction to bifurcation
theory.
\newblock {\em Rev.\ Mod.\ Phys.} {\bf 63-64} 991 (1991). 

\bibitem{col03} A.A. Coley.
\newblock {\em Dynamical systems and cosmology}.
\newblock (Kluwer Academic Publishers 2003).

\bibitem{uggetal03} C.~Uggla, H.~van Elst, J.~Wainwright, and
    G.~F.~R.~Ellis. The past attractor in inhomogeneous
    cosmology.
\newblock {\em Phys.\ Rev.} D {\bf 68} 103502 (2003).

\bibitem{heietal07} J.M. Heinzle, C. Uggla, and N. R\"ohr.
\newblock The cosmological billiard attractor.
\newblock To appear in {\em Adv.\ Theor.\ Math.\ Phys.} (2009). 
Electronic Preprint: arXiv:gr-qc/0702141.

\bibitem{heiuggpreprint} J.M. Heinzle and C. Uggla.
\newblock Mixmaster: Fact and Belief.
\newblock Preprint (2009).





\bibitem{janugg99}
R.T. Jantzen and C. Uggla.
\newblock The kinematical role of automorphisms in the orthonormal
frame approach to Bianchi cosmology.
\newblock  {\it J.\ Math.\ Phys.} {\bf 40} 353 (1999).

\bibitem{friren00}
H. Friedrich, A.D. Rendall.
\newblock The Cauchy Problem for the Einstein Equations.
\newblock {\it Lect.\ Notes\ Phys.} {\bf 540} 127 (2000).

\bibitem{andren01}
L. Andersson, A.D. Rendall.
\newblock Quiescent cosmological singularities
\newblock {\it Commun.\ Math.\ Phys.} {\bf 218} 479 (2001). 

\bibitem{linwal89}
X-f. Lin and R.M. Wald.
\newblock Proof of the closed-universe-recollapse conjecture for
diagonal Bianchi type IX cosmologies.
\newblock {\it Phys.\ Rev.\ D} {\bf 40} 3280 (2003).

\bibitem{heietal05}
J.M. Heinzle, N. Rohr, and C. Uggla.
\newblock Matter and dynamics in closed cosmologies.
\newblock {\it Phys.\ Rev.\ D} {\bf 71} 083506 (2005).

\bibitem{jan01}
R.T. Jantzen.
\newblock Spatially Homogeneous Dynamics: A Unified
Picture.
\newblock in {\it Proc.\ Int.\ Sch.\ Phys.\ ``E. Fermi" Course LXXXVI on
``Gamov Cosmology"\/}, R. Ruffini, F. Melchiorri, Eds. (North
Holland, Amsterdam, 1987) and in  {\it Cosmology of the Early
Universe\/}, R. Ruffini, L.Z. Fang, Eds. (World Scientific,
Singapore, 1984).







\bibitem{heiugg06}
J.M. Heinzle and C. Uggla.
\newblock Dynamics of the spatially homogeneous Bianchi type I Einstein-Vlasov
equations.
\newblock {\it Class.\ Quantum Grav.} {\bf 23} 3463 (2006). 

\bibitem{calhei07}
S. Calogero and J.M. Heinzle.
\newblock  Dynamics of Bianchi type I elastic spacetimes.
\newblock {\it Class.\ Quantum Grav.} {\bf 24} 5173 (2007).

\bibitem{colste71}
C.B. Collins and J.M. Stewart.
\newblock Qualitative Cosmology.
\newblock {\it Mon.\ Not.\ R.\ Astron.\ Soc.} {\bf 153} 419 (1971).

\bibitem{rin03}
H. Ringstr\"om.
\newblock Future asymptotic expansions of Bianchi VIII vacuum metrics.
\newblock {\it Class.\ Quantum Grav.} {\bf 20} 1943 (2003). 

\bibitem{waietal99}
J. Wainwright, M. J. Hancock,and  C. Uggla.
\newblock Asymptotic self-similarity breaking at late times in cosmology.
\newblock {\it Class. Quant. Grav.} {\bf 16} 2577 (1999). 

\bibitem{limetal06}
W.C. Lim, C. Uggla, and J. Wainwright.
\newblock Asymptotic silence-breaking singularities.
\newblock {\it Class. Quant. Grav.} {\bf 23} 2607 (2006). 

\bibitem{waihsu89}
J. Wainwright and L. Hsu.
\newblock A dynamical systems approach to Bianchi cosmologies:
orthogonal models of class A.
\newblock {\it Class.\ Quantum Grav.} {\bf 6} 1409 (1989). 









\end{thebibliography}

\vfill

\end{document}